\documentstyle[emulateapj,epsfig]{article}
\renewcommand{\baselinestretch}{1.0}
\received{}
\revised{}
\accepted{}
\journalid{}{} 
\articleid{}{}
\paperid{}
\cpright{}{}
\ccc{}
\lefthead{Uson \& Matthews}
\righthead{{H\,{\sc i}} Imaging of UGC~7321}
\slugcomment{Submitted to the Astronomical Journal}
\begin{document}

\vglue0.6truein

\title{{H\,{\sc i}}  Imaging Observations of Superthin Galaxies. I. UGC~7321}

\vskip0.2truein

\author{Juan M. Uson\altaffilmark{1}}
\author{L. D. Matthews\altaffilmark{2}}

\altaffiltext{1}{National Radio Astronomy Observatory, 520 Edgemont Road,
Charlottesville, VA 22903-2475 USA.  Electronic mail: juson@nrao.edu}

\altaffiltext{2}{Harvard-Smithsonian Center for Astrophysics,
60 Garden Street, MS-42, Cambridge, MA 02138 USA.  Electronic mail:
lmatthew@cfa.harvard.edu}

\newcommand{\ang}{\rm \AA}
\newcommand{\msun}{\mbox{${\cal M}_\odot$}}
\newcommand{\lsun}{\mbox{${\cal L}_\odot$}}
\newcommand{\kms}{\mbox{km s$^{-1}$}}
\newcommand{\HI}{\mbox{H\,{\sc i}}}
\newcommand{\HeI}{\mbox{He\,{\sc i}}}
\newcommand{\mhi}{\mbox{${\cal M}_{HI}$}}
\def\hst{{\it HST}}
\def\rsun{{\rm\,R_\odot}}
\def\etal{{\it et al.}}
\newcommand{\OIII}{\mbox{O\,{\sc iii}}}
\newcommand{\HII}{\mbox{H\,{\sc ii}}}
\newcommand{\NII}{\mbox{N\,{\sc ii}}}
\newcommand{\SII}{\mbox{S\,{\sc ii}}}
\newcommand{\nan}{Nan\c{c}ay}
\newcommand{\skp}{\mbox{  }\\}
\newcommand{\bmv}{\mbox{B--V}}
\newcommand{\bmr}{\mbox{B--R}}
\newcommand{\darkV}{$\frac{{\cal M}_{HI}}{L_{V}}$}
\newcommand{\dark}{${\cal M}_{HI}/L_{B}$}
\newcommand{\am}[2]{$#1'\,\hspace{-1.7mm}.\hspace{.0mm}#2$}
\newcommand{\as}[2]{$#1''\,\hspace{-1.7mm}.\hspace{.0mm}#2$}
\newcommand{\UV}{$u$-$v$}
\newcommand{\jks}{Jy km s$^{-1}$}

\begin{abstract}
We have used the Very Large Array to image the isolated ``superthin'' galaxy
UGC~7321 in the \HI\ line with a spatial resolution of 16$''$ and a spectral
resolution of 24~kHz (5.2~\kms).  We have reached  a sensitivity of
(0.36 -- 0.40)~mJy/beam per channel, which correspond to a column density
of (8 -- 9)$ \times 10^{18}$~atoms~cm$^{-2}$ ($1 \sigma$).  UGC~7321 has a
gas-rich disk with \mhi$= (1.06\pm 0.01) \times10^{9}$~d$_{10}^2$~\msun\
and \dark$= 1.0$ (d$_{10}$ is the distance to UGC~7321 in units of 10~Mpc,
the value adopted in this paper), and no detectable radio continuum emission
($F_{CONT} = 0.41 \pm 0.25$~mJy).
The global \HI\ distribution of UGC~7321 is rather symmetric and extends to
$\sim$1.5 times the optical radius ($D_{HI}=$ \am{8}{65}~$ \pm $~\am{0}{15}
at $n_{HI}=3 \times 10^{19}$ atoms cm$^{-2}$).  
An ``integral sign'' warp is observed in the \HI\ disk, commencing near the
edge of the stellar distribution, and twisting back toward the equatorial
plane in the outermost regions.  In addition, the position-velocity diagram
suggests the presence of a bar or inner arm within $\sim$40'' from the
center.
The rotation curve of UGC~7321 is slowly rising; it reaches its asymptotic
velocity of $\sim$110~\kms\ at $\sim$ \am {2}{5} from the center (about 0.9
optical radii) and declines near the edge of the \HI\ disk.  The ratio of the
inferred dynamical mass to the mass in gas and stars is
$\sim$12~d$_{10}^{-1}$, implying that UGC~7321 is a highly dark-matter
dominated galaxy.
\end{abstract}


\section{Introduction\protect\label{intro}}
``Superthin'' galaxies appear ``needle-like'' on the sky because of their
edge-on orientation, highly flattened stellar disks ($a/b\ga$10) and (near)
absence of a discernable spheroidal component (Vorontsov-Vel'yaminov~1967).
A spectrophotometric study showed them to have gas-rich, optically diffuse
disks with little internal absorption, as well as low emission-line
intensity ratios and slowly rising rotation curves (Goad \& Roberts~1981).
Therefore, many of the so-called superthins are late-type, low surface
brightness (LSB) disk galaxies seen at high inclination (Matthews,
Gallagher, \& van~Driel~1999; Dalcanton \& Bernstein~2000; see also
Gerritsen \& de~Blok~1999).  They seem to be among the least evolved disk
galaxies in the local universe, having undergone only minimal dynamical
heating, star formation, and angular momentum transport (Bergvall \&
R\"onnback 1995; Matthews et al.~1999; Dalcanton \& Bernstein~2000).  In
addition, their orientation and simple structure make superthins ideal
laboratories for probing how galaxy disk structure and morphology are
influenced by internal and external processes.

Observational and theoretical arguments suggest that disk structure and
thickness are affected by environment (T\'oth \& Ostriker~1992;
Odewahn~1994; Reshetnikov \& Combes~1997; Schwarzkopf \& Dettmar~2001) and
imply that a disk must remain isolated in order to remain ``superthin.''
This is often observed, as most galaxies with $a/b\ge10$ are devoid of close
neighbors (Karachentsev 1999; Matthews \& van~Driel~2000).  However, some
exceptions do exist (Duc et al.~2000; Hibbard et al.~2001; Matthews \&
Uson~2002).

Since superthin galaxies are rich in neutral hydrogen (\HI) gas (Giovanelli,
Avera, \& Karachentsev~1997; Matthews \& van~Driel~2000), imaging
observations of the \HI\ (21-cm) line emission can be used to test whether
these objects have truly remained isolated throughout their lifetimes.
Dynamical timescales of the outer regions of the gaseous disks of galaxies
are longer than those of their stellar disks;  hence signatures of past
perturbations and interactions such as warps and gas asymmetries can be
preserved in the \HI\ distribution, even when none are obvious in the stellar
disk (Rix \& Zaritsky~1995).  In addition, high-sensitivity \HI\ observations
can uncover the presence of previously unseen gas-rich companions which (if
present) might also play a role in the evolution of these disks.

We can gain further insight into the nature of superthin galaxies by
comparing \HI\ images with data obtained at other wavelengths. For example,
measurements of the thickness of the stellar and gaseous disks of superthins
are of considerable interest in understanding the evolution of these
galaxies.  While the stellar scale-heights of superthin disks are typically
found to be quite small ($\la$200 pc, Matthews 2000; but see also
van~der~Kruit et al.~2001), this does not necessarily mean that their gas
disks are also unusually thin, since gaseous and stellar disks are affected
by different heating processes.  Comparison of the structure and scale
heights of the gas disks of superthins with those of other types of galaxies 
should constrain how the dynamically coldest disks are stabilized, and what
mechanisms are responsible for heating and supporting their ISM component,
especially in the absence of significant energy input from massive star
formation (Sellwood \& Balbus~1999; S\'anchez-Salcedo~2001).

Since the stellar disks of these galaxies remain so thin, internal as well as
external heating processes must be suppressed in them, including `firehose'
instabilities (Fridman \& Polyachenko~1984), bar formation (Mihos et
al.~1997), and the development of spiral arms (Noguchi 1987).  In addition,
their stability requires the presence of a massive dark halo (Efstathiou,
Lake and Negroponte~1982).  In fact, superthins should be amongst the most
dark-matter dominated of disk galaxies (Zasov, Makarov, \& Mikhailova~1991;
Gerritsen \& de~Blok~1999).   However, constraints on the dark matter content
of edge-on superthin disks have been derived for only a few objects such as
NGC~4244 (Olling~1996).  Existing optical rotation curve studies of
superthins typically show a slow rise throughout much of the stellar disk
(e.g., Goad \& Roberts~1981; Makarov et al.~1997);  but only \HI\
observations permit the derivation of the extended outer disk rotation curves
which are necessary to constrain the mass models and dark matter halo
parameters.  For superthins, such analyses are not complicated by the
presence of a bulge component, and the mass contributions of stellar disks
are expected to be low, making them particularly well-suited for probing the
shape of the distribution of dark matter in these galaxies (Olling 1995).

In spite of the unique information that can be gleaned from detailed studies
of edge-on spiral galaxies, only a small number of these have been imaged in
the \HI\ line with high sensitivity and high spatial and spectral resolution
(e.g., Rupen 1991;  Olling 1996;  Swaters, Sancisi, \& van der Hulst~1997;
Garc\'\i a-Ruiz, Sancisi, \& Kuijken~2002).  Moreover, with the exception of
Olling (1996), the most sensitive \HI\ imaging observations of edge-on
galaxies to date have concentrated on massive, luminous systems of
mid-to-early Hubble type rather than on moderate-to-low mass and/or LSB
systems like the superthins.

This paper is the first in a series describing new high-sensitivity \HI\
imaging observations of superthin galaxies with moderate optical luminosities
($L_{B}\sim10^{9}$~\lsun), moderate masses ($V_{ROT}\sim$100~\kms), and
highly flattened disks ($a/b\sim$10).  The galaxies that we have observed are
undergoing differing levels of perturbation from a companion, ranging from
being isolated to being strongly perturbed, and show differing levels of
current star formation.
In this paper, we present our study of UGC~7321, an isolated superthin galaxy
which offers a unique perspective on the ISM structure of an LSB disk viewed
edge-on.  By comparing its \HI\ properties to those of other LSB and edge-on
spiral galaxies, and ultimately, to the results to be presented in our
following papers, we aim to address many of the questions discussed
above---i.e., to gain clues on the formation histories of pure disk galaxies
and probe how environmental factors, interactions, and internal processes
influence the structure, evolution, and star formation histories of late-type
disk galaxies.
\vskip 0.35truein

\section{The Target: UGC~7321\protect\label{target}}
UGC~7321 is a highly inclined ($i = 88 \pm 1^{\circ}$) Sd spiral galaxy.
Its distance is somewhat uncertain, with estimates ranging from $\sim$5~Mpc
from a kinematical model of the Local Supercluster (Tully, Shaya, \&
Pierce~1992) to 14.9~Mpc using the $B$-band Tully-Fisher relation
(Garc\'\i a-Ruiz et al.~2002).  One of us (LDM) has used the {\it Hubble
Space Telescope} to obtain images of UGC~7321 in collaboration with
J.~S.~Gallagher (Wisconsin) for a different project.  They estimate that
UGC~7321 is located at a distance in the range of (7 -- 13)~Mpc based on the
brightest resolved stars.  In this paper, we adopt a distance of 10~Mpc for
the derivation of physical parameters, and indicate the dependence of our
results with distance explicitly by using the symbol d$_{10}$ to denote
distance in units of 10~Mpc.  We summarize the optical and near-infrared
properties of UGC~7321 in Table~1.

The scale height of the disk of UGC~7321 is one of the smallest ever reported
for a galaxy disk (Matthews~2000).  UGC~7321 is also unique among the
well-studied edge-on spirals because it has an extremely diffuse stellar disk
with only minimal patchy dust obscuration and very low internal absorption.
Correspondingly, UGC~7321 has very low far-infrared and CO luminosities
(Matthews \& Gao~2001).

UGC~7321 is a very \HI-rich galaxy with a high \dark\ ratio. Indeed, Haynes
et al.~(1998) obtained a carefully calibrated (5\%) \HI\ spectrum of UGC~7321
with the Green Bank 43~m telescope, which yielded an integrated \HI\ line
flux density of 44.51~Jy~km~s$^{-1}$.  Matthews et al.~(1999) made a
7-pointing map with the \nan\ decimetric radio telescope and showed that the
\HI\ gas extends beyond 1.2~times the optical radius.  We compare these
single-dish observations with our new VLA \HI\ measurements in more detail in
Section~\ref{global}.  It has also been included in the recent study of warps
in edge-on galaxies of Garc\'\i a-Ruiz et al.~(2002) who used observations
made for the WHISP survey.
\vfil\eject

\section{Observations\protect\label{obs}}
We observed UGC~7321 with the Very Large Array (VLA)\footnote{The Very
Large Array of the National Radio Astronomy Observatory is a facility
of the National Science Foundation, operated under cooperative agreement
by Associated Universities, Inc.} in C configuration during 2000 May 26
(hereafter Day~1) and 2000 May 30 (Day~2).  Since early 1998, the
C~configuration is a modified version of the original C~configuration
(3~km maximum baseline) in which the fifth antenna of the North arm is
placed at the center of the array thus improving its sensitivity to extended
emission.  Indeed, it allowed us to ``recover'' (nearly) all of the
``single-dish'' \HI\ signal from UGC~7321 (see Section\ref{global}).  The
observations were obtained during two 8-hour (late-afternoon to mostly
nighttime) observing sessions.

To obtain the best possible velocity resolution over the frequency range of
interest (defined from the single-dish \HI\ observations), we observed the
field centered at the optical position of the galaxy with spectral mode~`4'
using two pairs of 127~channels with on-line Hanning smoothing; this
produced two pairs of 63~nearly-independent, contiguous channels of width
24.4~kHz ($\sim$5.2~\kms).  For both observing sessions, the first
``IF pair'' (consisting of separate right and left circular polarizations)
was tuned to the known heliocentric velocity of UGC~7321 (407~\kms), while
the second ``IF pair'' was tuned to a central velocity of 650~\kms (on
Day~1), and 300~\kms (on Day~2) in order to search for possible dwarf
companions or high-velocity gas in the neighborhood of UGC~7321 (see
Section~\ref{COMPANIONS}).  Observations of the point source 1221+282
(5~minute ``scans'' each) were interspersed with observations of UGC~7321
(30~minute scans).  The strong source 1331+305 (3C286) was observed twice
each day (10~minute scans) in order to obtain good flux calibration.  Most of
the data on UGC~7321 (720~minutes) were obtained at elevations above
40$^\circ$.  Details appear in Table~2.

\section {Data Reduction\protect\label{reduce}}
\subsection{Calibration\protect\label{calib}}
Because our chosen phase-calibrator (1221+282) turned out to be considerably
weaker (0.86~Jy) than the value given in the VLA calibrator list (1.8~Jy)
we used a non-conventional path for the data reduction.  We generated a
pseudo-continuum database by vector-averaging the data for channels~4
through~60 at each time stamp, but only used this file to find and edit
corrupted data (mostly due to interference and cross-talk at small spacings
as well as a few malfunctioning correlators).

We determined the overall shape of the bandpass as well as the zeroth-order
amplitude and phase calibration in one operation using the strong primary
calibrator 1331+305 (3C286, which was observed at an elevation intermediate
to those corresponding to the observations of UGC~7321).  We adopted a flux
of 14.7363~Jy at a frequency of 1418.36~MHz, appropriately scaled to the
other two frequency settings, as indicated in the VLA calibration manual.
This bandpass calibration was applied to all data, which provided the
amplitude scale of each frequency channel as well as an initial calibration
of the instrumental phases.  As is well known, phase errors increase at the
band-edges and prevent an accurate calibration of the affected channels.
We found that channels 1--3 and 61--63 could not be calibrated to the
accuracy of the 57~central channels.  Therefore, we have not used those six
edge-channels in the analysis that follows.

Next, we made a (deconvolved) continuum image of the phase-calibrator using
the spectral data from channels~4 through~60 with each visibility gridded at
the correct ``uv-cell.''  This image was used for a subsequent ``global''
amplitude and phase (self) calibration (i.e., one amplitude and phase
correction per antenna per scan which was applied to all frequency channels)
in which we constrained the average of the amplitudes to remain constant (a
global re-scaling by the factor 0.996) and where the averaging time was that
of a ``scan.''  These corrections were subsequently applied to the UGC~7321
visibilities after linear 2-point interpolation (in time) of amplitudes and
phases.

These steps completed the calibration of the observations of Day~1, as we
obtained two sets of spectral images (``image cubes'') that were free of any
noticeable artifacts.  In addition, the spectrum of the strongest continuum
source in the UGC~7321 field (J1217+2239, with an observed flux density of
64.5~mJy), located at an angular distance of \am{8}{6}, did not show any
systematic effects and was well-fitted with a constant over both frequency
settings.

On the second day, the VLA was not quite as stable as on the first day and we
found it necessary to improve on the calibration of the instrumental
antenna-based passbands.  We followed the operations described above with a
secondary calibration of the passband determined with the data obtained on
J1221+282 (the phase calibrator).  Because of the low flux density of this
calibrator (0.86~Jy) and the confusion in the surrounding area ($\sim$0.3~Jy)
we could only afford to determine one (complex) correction per antenna and
per channel for the whole observing run, which was obtained by comparing the
visibilities channel-by-channel with the theoretical values obtained from the
continuum image of J1221+282 obtained as described above.  These corrections
were subsequently applied to the UGC7321 data and new image cubes were made.
This step was sufficient for the second frequency setting used on this day,
but not so for the first frequency setting, as the spectrum through the
aforementioned continuum source, J1217+2239, showed significant curvature.
Therefore, we were forced to evaluate scan-based corrections to the spectral
passband for the main frequency-setting on this day using the observations of
J1221+282 (the phase calibrator).  However, given the low signal-to-noise
ratio per channel of these observations, we could only determine (scan based)
global corrections to the passband in the form of 5-th order Chebyshev
polynomials using channels~4 through~60 by comparing the so-corrected data to
the theoretical visibility function derived from the continuum image of the
J1221+282 field determined in the previous step.  These corrections were once
again applied to the UGC7321 data using linear-interpolation between pairs of
observations of 1221+282.  The image cube determined after this calibration
was free of defects, and the source J1217+2239 discussed above was well fit
with a constant.

We have performed a number of tests on the continuum images of our amplitude
and phase calibrators and estimate that the uncertainty in the amplitude
scale of the images that we present in this paper is about 1\%.
\subsection{Imaging of the Continuum Emission\protect\label{imagcont}}
We used the line-free channels in the second frequency setting (hereafter
IF2) of each observation to compute an image of the continuum emission in
the vicinity of UGC7321.  Because there were 47~such channels on Day~1 and
only 23~on Day~2, we proceeded in the following manner.  We first made a
multi-field (20 fields) image with the data from channels 4--50 from Day~1
using channel-based ``3-D'' gridding and a robustness factor $r = 0.7$,
which offers an optimal compromise between noise and resolution when imaging
VLA observations obtained with full UV-coverage (Briggs 1995).  This yielded
a continuum image computed over an effective bandwidth of $\sim$1.15~MHz.
These images were used to generate phase corrections through self-calibration
of the data with a 5~minute averaging interval.  Next we subtracted the
theoretical visibilities derived from the images on a channel-by-channel
basis and subsequently determined the frequency-averaged residual
visibilities.  The increased sensitivity of these pseudo-continuum
visibilities allowed us to recognize some residual low-level contamination
which was mostly due to solar interference (in the data obtained in the
late-afternoon) as well as cross-talk on short spacings.  These data were
discarded (about 11\% of the visibilities) prior to restoring the
(theoretical) contribution of the radio sources determined above.  The same
procedure was applied to the line-free channels (38--60) in IF2 of Day~2,
except that here a 10~minute averaging time was used for the phase-only
self-calibration because of the reduced effective bandwidth
($\sim 0.56$~MHz).  Again, about 10\% of the visibilities were found to
show low-level contamination and discarded.  The resulting pseudo-continuum
visibilities were combined and a new set of final images (20~fields) was made
using a $2''$ cell-size.  The images have a nearly circular synthesized-beam
with $FWHM \sim 15''$ and an rms noise level of $66 \mu$Jy/beam
(essentially the theoretical noise level), and are devoid of any noticeable
artifacts.  The total deconvolved flux is 199~mJy.

\subsection{Imaging of the Line Emission\protect\label{imagline}}

It is customary to remove the continuum emission from the visibilities prior
to imaging the line emission in the individual channels.  This is often
accomplished with a visibility-based subtraction of the continuum in which
a first-order polynomial is fit to the real and imaginary components of each
visibility, and the residuals are used to make the channel images (Cornwell,
Uson, \& Haddad~1992).  In our case, there are only four line-free channels
at each end of IF1 (channels 4--7 and 57--60) and the limited number of
degrees of freedom in the visibility-based subtraction of the continuum
results in an artificial lowering of the noise in the images corresponding
to those channels (to $\sim$75\% of the theoretical level).  Therefore, we
have followed an alternate path to compute the channel images, whereby we
have made images of the total (line plus continuum) emission in each channel
and subsequently determined the line emission as discussed below.

We made twelve image cubes using the combined data from both of our observing
dates for IF1.  Only the first image cube (which is centered on UGC~7321)
contained \HI\ emission; the remaining eleven image cubes were centered on
neighboring continuum sources in order to image the total continuum flux in
the field properly.  We used a $3''$-cellsize and a robustness factor
$r = 1$ in order to minimize the noise level while still yielding a
reasonable synthesized-beam which was nearly circular with
$FWHM = 16.2'' \times 15.8''$ at position angle PA$ = -34^\circ$.  The
deconvolution was deep, to 0.35~mJy~beam$^{-1}$ ($\sim 1 \sigma$) and was
followed by a restoration of spurious (isolated) components if the absolute
value of the flux inside a six-pixel radius was below 0.7~mJy~ beam$^{-1}$.
The peak deconvolved flux (line plus continuum, uncorrected for primary-beam
attenuation) was 480~mJy at 1417.922~MHz.  The rms noise was
0.35--0.40~mJy~beam$^{-1}$.  This procedure resulted in images that were free
of artifacts and did not show a negative bowl surrounding the \HI\ emission.

We made the corresponding sets of twelve image cubes with the data from IF2
for each one of the two observing dates.  Because the noise level was
40\% higher than for the combined data from IF1, we stopped the deconvolution
and ``filtered'' the spurious components at correspondingly scaled levels.

Next we combined the three image cubes corresponding to the central field at
the three observed frequency settings into a global cube with 125~channels in
the following way:  Channels~1 through~47 correspond to channels~4 through~50
of IF2 from Day~1.  Because the IF settings used
on Day~1 resulted in overlapping channels that were reasonably aligned (a
misalignment of only $\sim$4\%) channels~48 through~57 are a weighted
average of the images in channels~51 through~60 of IF2 (Day~1) and channels~4
through~13 of IF1 (both days) respectively.  Most of the weight is attached
to the second image in each pair (created from the combined data from both
days).  This might seem unnecessary and even undesirable, as corresponding
channels observed through both IFs (on Day~1) view the same sky-photons, but
the VLA backend introduces some noise into the observation which can be
reduced somewhat through this average.  Care was taken that the weights
accurately reflected the loss of independence of the images averaged.
Channels~58 through~101 of the global cube were channels~14 through~57 of
IF1 (obtained from the combined observations of both days); while
channels~102 through~104 resulted from the weighted average of channels~58
through~60 of IF1 and~37 through~39 of IF2, Day~2, respectively, with
weights determined as above.  Because the frequency alignment of the
channels was somewhat poorer on Day~2 (a misalignment of $\sim$25\%) here
we have only combined line-free channels, so that the small blurring (in
frequency) is of no consequence.  Finally, channels~105 through~125 of the
global cube correspond to channels~40 through~60 of IF2 of Day~2.

\begin{figure*}
\plotone{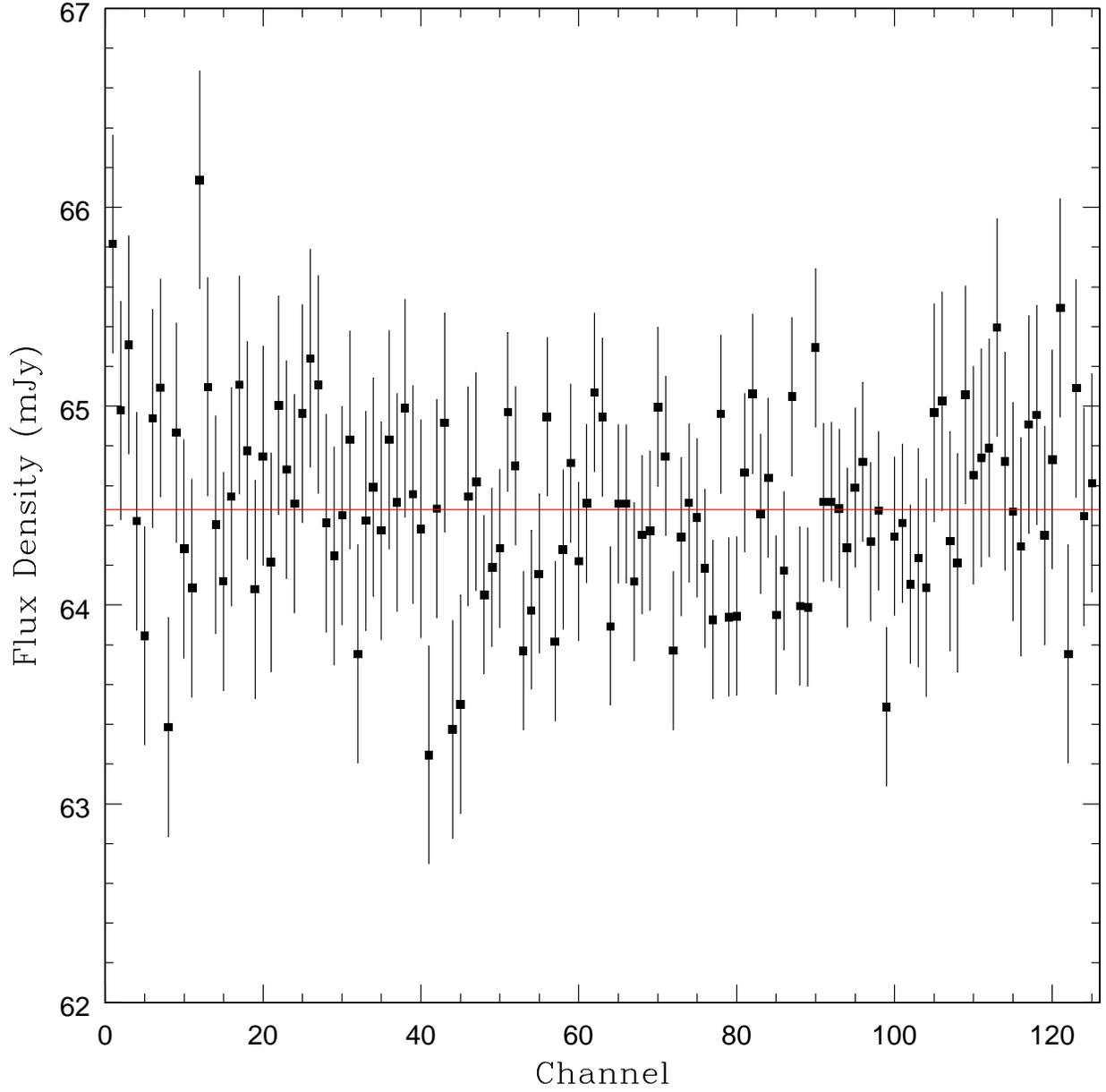}
\figcaption{
Observed spectrum of J1217+2239, the strongest continuum source
in the vicinity of UGC~7321.  The rms noise is $\sim 0.54$~mJy for channels
1--47 and 103--125, and $\sim 0.38$~mJy for channels 48--102.  The data are
well fit by a constant value of S~$= 64.48 \pm 0.04$~mJy (solid line).  No
correction for primary-beam attenuation has been applied.
\protect\label{fig:contspectrum}}
\end{figure*}

We have performed a number of tests on this combined image cube.
Figure~\ref{fig:contspectrum} shows the spectrum at the position of the
continuum source J1217+2239 in the combined, global 125-channel cube.  It
is well fit by a constant  $S = 64.48 \pm 0.04$~mJy with a Chi-square of
123.9 with 124 degrees of freedom.  This is an indication that the bandpass
correction has been successful in flattening the spectrum and that the
amplitude and phase calibrations on the four different IF-Day combinations
are consistent to within the noise level reached.  It also shows that the
deconvolution has been successful in eliminating the bowl around the \HI\
emission that is quite apparent in the raw images and that any residual
deconvolution bias is fairly consistent from channel to channel.

\begin{figure*}
\plotone{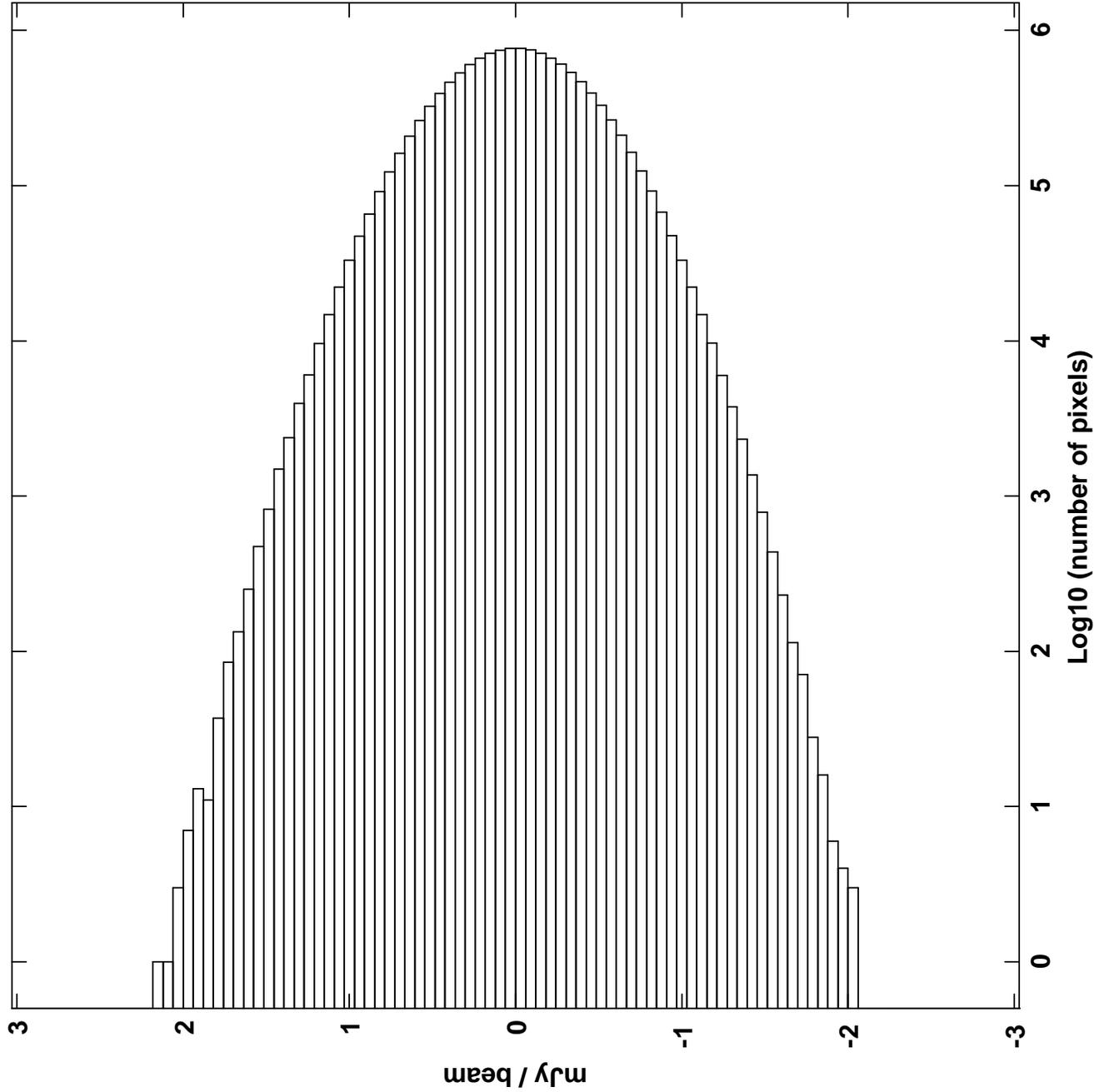}
\figcaption{
Histogram showing the noise distribution in the image-planes of
UGC~7321 corresponding to the IF1 setting.  These are planes 48--104.  The
central region ($9' \times 3'$ for planes 52--100) which contains the \HI\
emission signal has been excluded.
\protect\label{fig:imeannoise}}
\end{figure*}

\begin{figure*}
\plotone{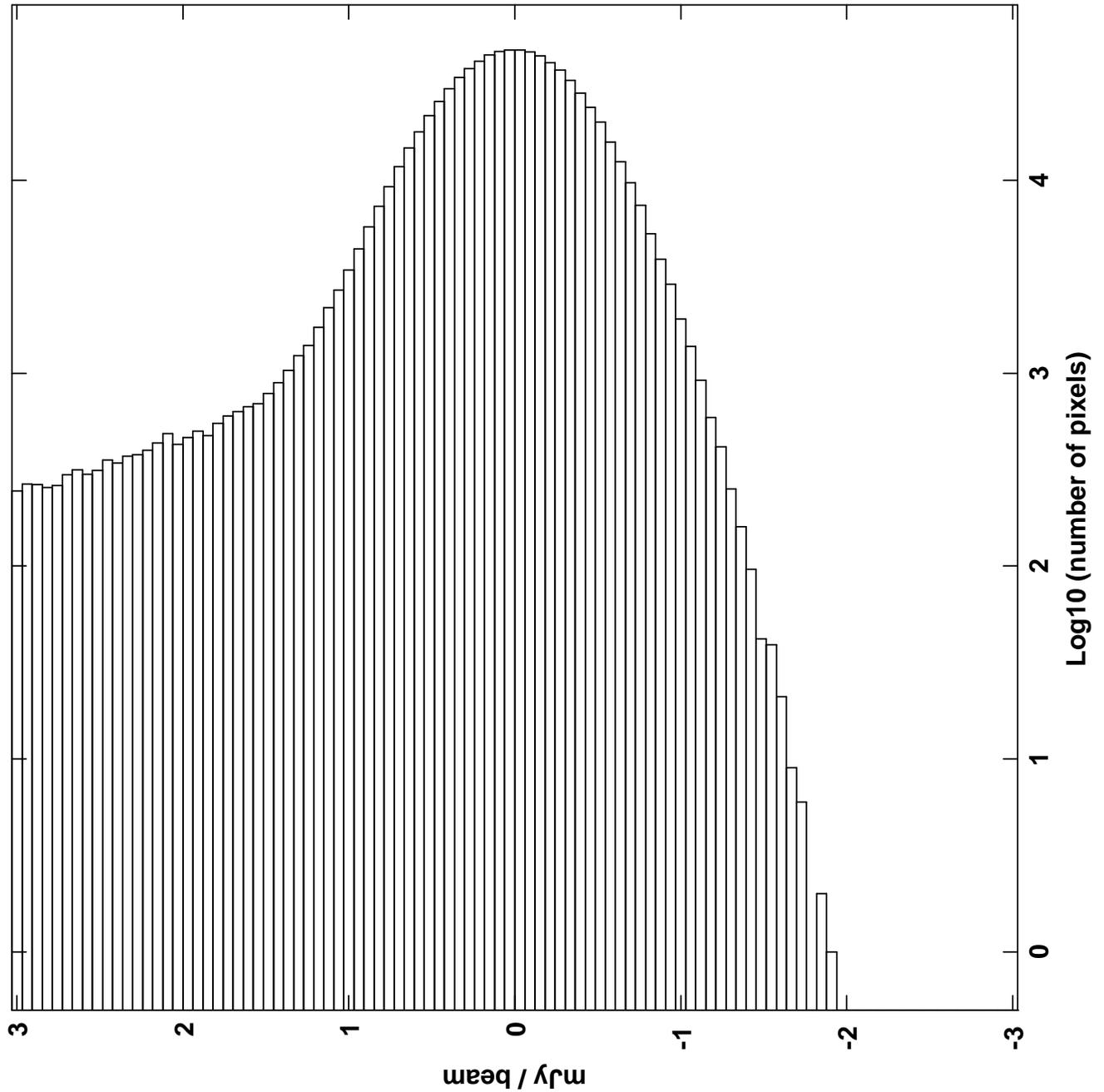}
\figcaption{Histogram showing the distribution of flux in the central region
($9' \times 3'$ for planes 52--100) which contains the observed \HI\
emission.  Notice the positive tail due to the signal; but also that the
negative values show the distribution that is expected in the absence of
artifacts.
\protect\label{fig:imeansignal}}
\end{figure*}

We have examined the statistics of the image cube to check for artifacts.
The cube is remarkably clean.  For example, Figure~\ref{fig:imeannoise}
shows a histogram of the values in planes 48--104 which correspond to IF1
but with the region of \HI\ emission from UGC~7321 excluded (the central
$9' \times 3'$ for planes 52--100).  The histogram is as expected from a
statistical distribution with the corresponding number of degrees of freedom
(values are correlated within any synthesized-beam area).  Even more
remarkable is the comparison with the histogram in
Figure~\ref{fig:imeansignal}, which shows the distribution of values in the
central $9' \times 3'$ for planes 52--100 that had been excluded from
Figure~\ref{fig:imeannoise}.  The signal is clearly noticeable, but it is
also clear that the negative values show the distribution that is expected
in the absence of any deconvolution artifacts, such as bowls.

We have determined the continuum emission from this cube in the image-plane
by linear fitting of a first-order polynomial to each line-of-sight (pixel)
in the cube (Cornwell et al.~1992) after blanking the region that contains
the \HI\ emission from UGC~7321.  Because of the finite spacial distribution
of the \HI\ emission, this procedure results in a larger number of
``line-free'' channels per line-of-sight than if the subtraction is performed
in the visibility domain and the procedure does not alter the noise
characteristics of the channel images in a detectable way.  This
continuum-subtracted cube has been used for the analysis of the \HI\ emission
reported in this paper.

Finally, we have made a similar 125-channel image cube following the
procedure just described but with a robustness parameter $r = -1$.  This is
closer to uniform weigthing and results in a somewhat sharper synthesized
beam that is again nearly circular (FWHM$ \sim 12.4'' \times 12.0''$ at
PA~$ \sim -48^\circ$, with a $3''$ cellsize).  The weighting results in noise
levels that are about 40\% larger that the corresponding ones for the $r = 1$
cube and the deconvolution limits and filtering levels were correspondingly
scaled.

\section{Analysis}
\subsection{Continuum Emission\protect\label{cont}}
\begin{figure*}
\plotone{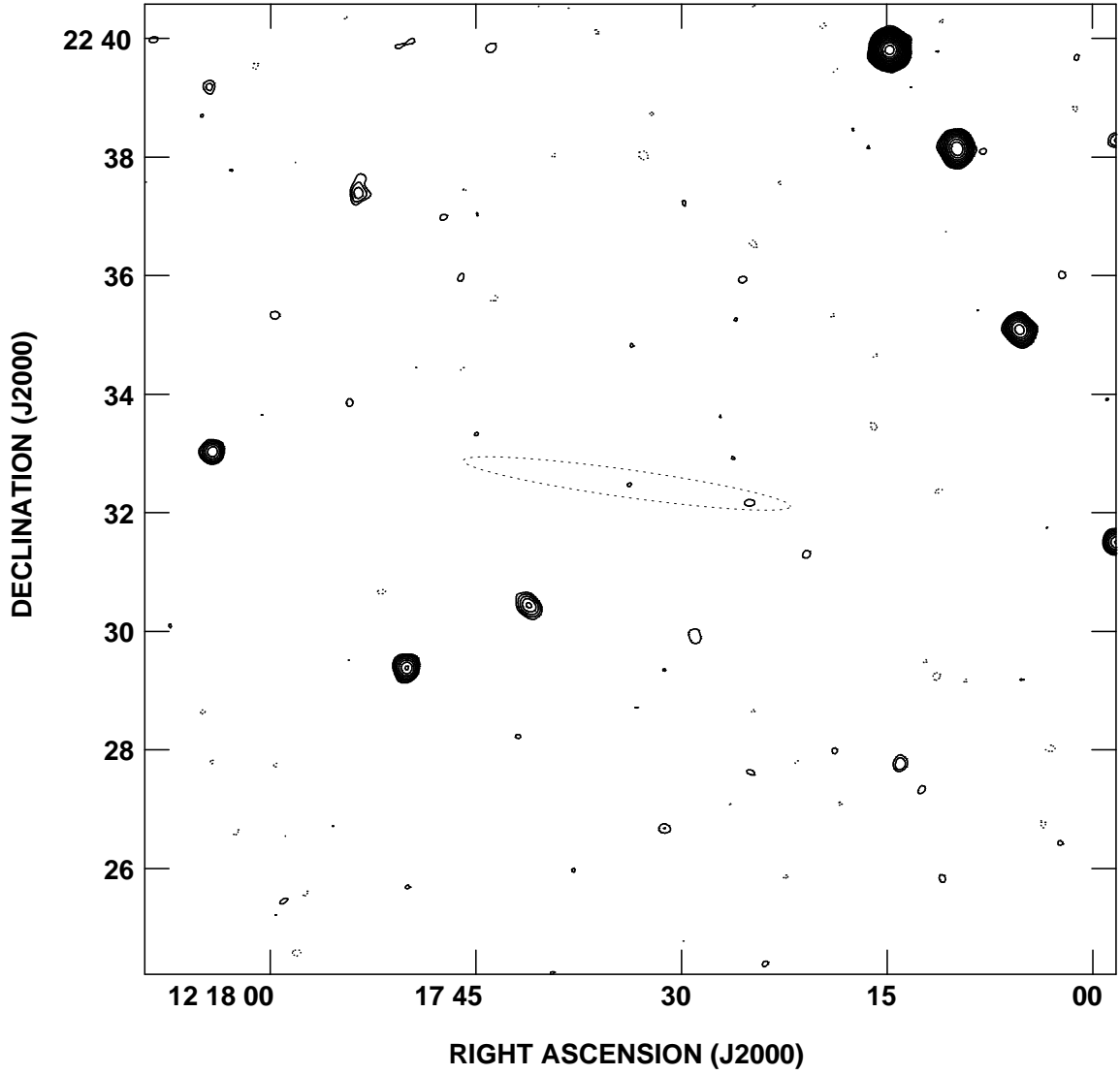}
\figcaption{Image of the continuum emission in the $16.4' \times 16.4'$
region centered on the position J1217+2232 at an effective frequency of
1418.0~MHz.  The dotted line shows the location of the edge of the optical
disk of UGC~7321.  Contour levels are [-2 [absent], -1.4, -1, 1, 1.4, 2,
2.8, 4, 5.6, 8, 11, 16, 22, 32, 44, 64, 88, 128, 176, 256,
352)$ \times 0.2$~mJy~beam$^{-1}$  (the $3 \sigma$~level)].
\protect\label{fig:continuum}}
\end{figure*}

The region around UGC~7321 contains only relatively weak radio continuum
sources (Figure~\ref{fig:continuum}).  The brightest source within the
primary beam, J1217+2239, is located $8.6'$ to the NW of UGC~7321 and has an
integrated flux of $81.5 \pm 0.2$~mJy (after correcting for the attenuation
due to the primary beam), in good agreement with the value $83.9 \pm 2.5$~mJy
listed in the NRAO VLA Sky Survey (hereafter NVSS; Condon et al.~1998).

It is clear from Figure~\ref{fig:continuum} that we have failed to detect
continuum emission from UGC~7321 itself.  The peak value within the (optical)
disk of the galaxy is only $S = 0.2 \pm 0.07$~mJy~beam$^{-1}$ (formal error)
at 12$^h$ 17$^m$ 33.8$^s$ +$22^\circ 32' 29''$(J2000).  However, because of
the uncertainty introduced by the deconvolution of the sidelobes from
J1217+2239, the actual uncertainty at this position is somewhat larger.

UGC~7321 was only weakly detected in the FIR by {\it IRAS} in the
60$\mu$m and 100$\mu$m bands.  Upon examining the {\it IRAS} 60$\mu$m and
100$\mu$m survey scans from the NASA/IPAC Infrared Science Archive, we found
UGC~7321 to be consistent with a point source in the FIR with FWHM$\sim1'$.
The {\it IRAS} faint-source catalog indicates flux densities  of 0.344~Jy at
60$\mu$m ($9 \sigma$) and 0.964~Jy at 100$\mu$m ($5 \sigma$) which yields a
FIR flux of $2.3 \times 10^{-14}$~W m$^{-2}$ (using eq.~14 in Condon 1992)
and a predicted 1.4~GHz continuum flux density of $\sim$3.1~mJy (eq.~15 in
Condon 1992).  To place a meaningful limit on the radio continuum emission
from UGC~7321, we have summed the flux in our continuum image
(Section~\ref{imagcont}) within a $1' \times 1'$ box centered at the
coordinates quoted above, which yields a flux density of $0.41 \pm 0.25$~mJy
Given the low significance of the {\it IRAS} detections, it is perhaps not
surprising that the galaxy is not detected in the radio continuum.  In
addition, some of the FIR luminosity could be due to a contribution from
cirrus emission as well as from a weak embedded nucleus.  Indeed, Matthews et
al.~(1999) found a rather red ($B-R\approx$1.5), compact feature inside the
{\it IRAS} box.  Moreover, a low-mass galaxy such as UGC~7321 might be unable
to confine cosmic rays with sufficient efficiency and would thus fall short
of the prediction from the FIR-radio correlation.

Using the relations given by Condon~(1992), the low FIR emission from
UGC~7321 yields a rather small estimate of its high-mass star formation rate
of $\sim$0.006\msun/year (for $M_{\star}>5$\msun).  Such a rate is more than
one order of magnitude smaller than the mean value estimated for all UGC
galaxies of type Scd-Im (see Roberts \& Haynes 1994), confirming that
UGC~7321 is a bona fide low surface-brightness spiral galaxy.

\subsection{Channel Images}
\begin{figure*}
\plotfiddle{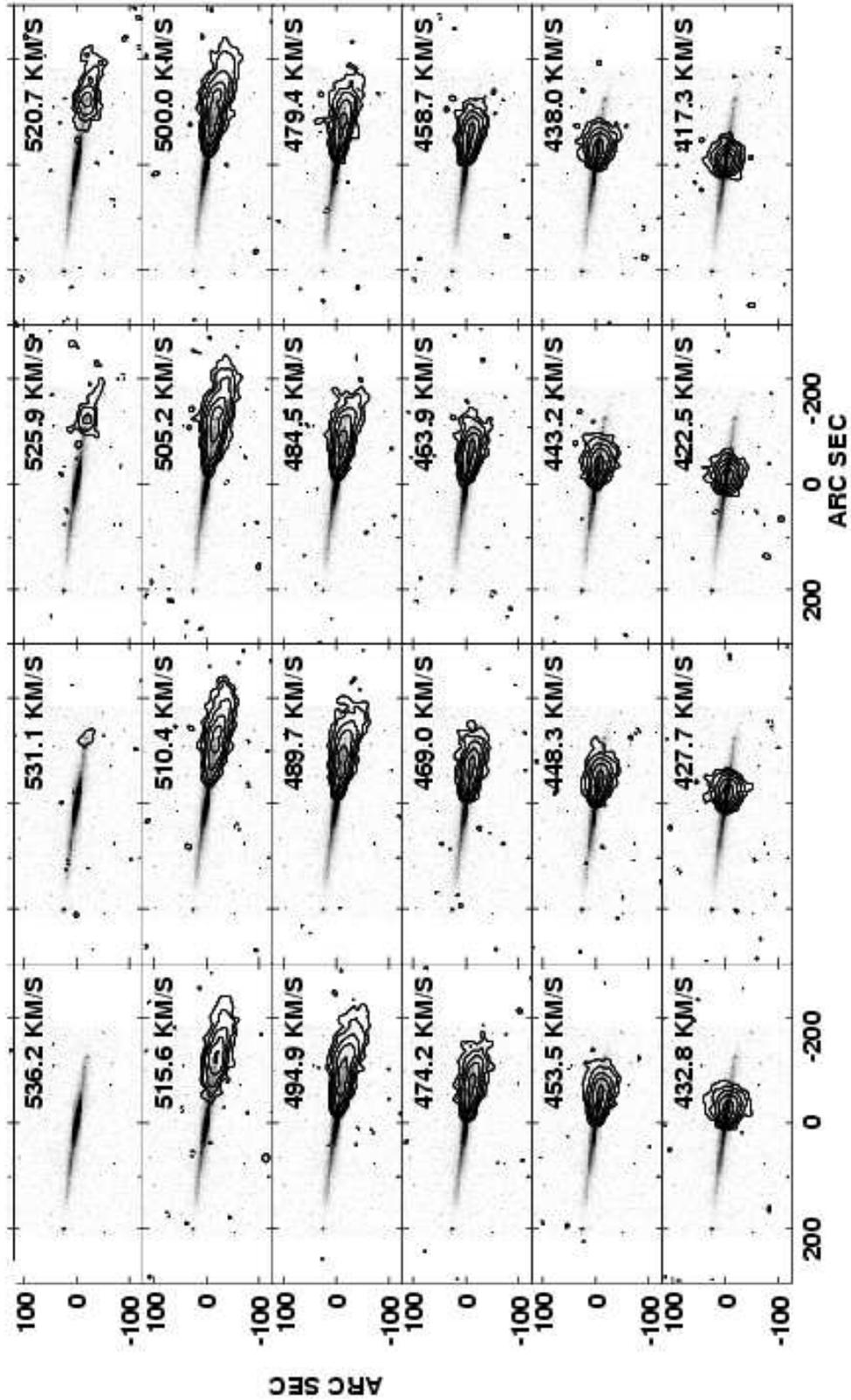}{22cm}{0}{85}{85}{-270}{-10}
\figcaption{\HI\ channel maps of UGC~7321, overplotted on optical $R$-band
images of the galaxy.  The maps have a spatial resolution of $\sim16''$ and
a velocity resolution of $\sim$5.2~\kms.  The contours are (-2 [absent], -1,
1, 2, 4, 8, 16, 32)$ \times 1.2$~mJy~beam$^{-1}$.  Only the channels showing
line emission (plus one additional channel on either side) are plotted.  Each
panel is labelled with its corresponding central heliocentric velocity.
\protect\label{fig:channelmaps}}
\end{figure*}

Figure~\ref{fig:channelmaps} shows the images corresponding to the channels
that contain detectable \HI\ emission, flanked with one extra channel on
each side, overlayed on an $R$-band image of UGC~7321.  A number of features
of the \HI\ disk and velocity field of UGC~7321 can be seen in these channel
images.  For example, the channels covering the velocity ranges 293.4 --
319.2~\kms\ and 479.4 -- 510.4~\kms, clearly show warping of the outer gas
layers.  In addition, in most of the channel images, the spread in $z$ of the
emission increases with distance from the disk center, which indicates
flaring of the gas layer.  However, the gas in the most extreme velocity
channels is confined to a narrower distribution than at small and
intermediate velocities, contrary to the expectation for a pure flaring model
(see also Swaters et al.~1997).  

It is interesting that in the two channels representing the velocity extrema
of the observed \HI\ emission (i.e., corresponding to velocities of 
283.0~\kms\ and 531.1~\kms), we see gas that is not located  at the edge of
the \HI\ disk, but rather at smaller galactocentric radii, near the edge of
the stellar disk.  This is indicative of a falling rotation curve in the
outskirts of UGC~7321 (see Section~\ref{rotcurve} below).

\subsection{Global Properties of the Neutral Hydrogen in UGC~7321}
\subsubsection{Total {H\,{\sc i}} Content \protect\label{global}}
In order to estimate the total \HI\ content of UGC~7321 and derive a global
\HI\ rotation profile, we measured the total \HI\ flux in each of the
channel images from our global, $r = 1$, cube, after discarding low values
(those with absolute value $\leq 0.5 \sigma$) and restricting the sum to
the pixels corresponding to the galaxy.  The corresponding pixels were
determined from an initial selection consisting of those pixels which were
above the $2 \sigma$ level (where the galaxy is quite distinct) and
expanding such areas by successive surrounding bands 3-pixels wide ($9''$)
until the enclosed flux converged.  For each channel, the selected areas
were summed after correction for the attenuation of the primary-beam.  We
estimated the error in the total \HI\ flux in each channel from the rms
noise in that channel, taking into account the correlation introduced by the
synthesized-beam and the correction due to the attenuation of the primary
beam.  The resulting errors are quite small ($\sim 3$~mJy per channel).  

The resulting global \HI\ profile is shown in Figure~\ref{fig:global}.  Also
shown are the profiles obtained with the NRAO 43~m~telescope (Haynes et
al.~1998) and with the~\nan\ telescope (Matthews et al.~1999).  The three
\HI\ profiles match extremely well given the 5\% calibration uncertainty 
(and statistical correction for primary beam attenuation) of the Green Bank
data, and the estimated calibration uncertainty of the seven-pointing \nan\
spectrum.  The total \HI\ flux derived from the present data is
$F_{HI} = 45.3 \pm 0.5$~Jy~km~s$^{-1}$, where the error comes from a
statistical contribution of 0.11~Jy~km~s$^{-1}$, an estimated systematic
contribution of 0.36~Jy~km~s$^{-1}$ and a global calibration uncertainty of
$\leq 1$\%.  Comparison with the values measured with the 43~m
(44.51~\jks\ ) and the \nan\ (47.6$\pm4.0$\jks\ with 15\% calibration
uncertainty) telescopes indicates that the measurements reported in this
paper have likely recovered most of the \HI\ flux.

The measured brightness temperatures peak at a value of T$_B = 97.8$~K at
the center of the galaxy and at the systemic velocity.  The somewhat better
resolution image made using a robustness factor r$ = -1 $ (FWHM$ \sim 12''$)
yields a corresponding peak brightness temperature T$_B = 118.3$~K.  This
value is interestingly close to those measured in the Milky Way (Burton~1970)
suggesting that the \HI\ spectrum of UGC~7321 might be affected by
self-absorption.  We have examined cuts along the major and minor axes of the
emission at the systemic frequency (which contains the peak values).  The
latter is well fit by a Gaussian of half-width $\sim24''$, with the data
falling short of the Gaussian fit at the center position by $\sim 2$\%,
indicating only mild self-absorption.  Neglecting this leads to an estimate
of the total \HI\ mass of UGC~7321 of \mhi\
$(1.06 \pm 0.01) \times 10^{9}$~d$_{10}^2$~\msun.

From the global \HI\ profile, we measure velocity widths at 20\% and 50\% of
the peak maximum of 234.3 and 219.8~\kms, respectively. The heliocentric
recessional velocity (optical definition) determined as the average of the
20\%-level values of the global \HI\ profile is 406.8~\kms (Table~3).  These
values are all in agreement with previous single-dish measurements.

\subsubsection{The Total {H\,{\sc i}} Intensity Distribution
\protect\label{totint}}
The sub-images containing \HI\ extracted as discussed in the previous section
have been added to obtain the total \HI\ intensity map of UGC~7321 which is
shown in Figure~\ref{fig:totintmap}, overplotted on a continuum-subtracted
H$\alpha + [\NII]$ image of the galaxy.  The overall \HI\ morphology of
UGC~7321 is typical of normal, edge-on spirals.  It shows a relatively
smooth, unperturbed, symmetric distribution, and we see no obvious
enhancements in the \HI\ emission corresponding to the locations of the
H$\alpha + [\NII]$ emission.  At a resolution of $\sim16''$, the peak
intensity observed within the galaxy corresponds to
$n_{HI}=7.3\times10^{21}$ cm$^{-2}$, while the maximum observed extent of
the \HI\ disk along the major axis is \am{8}{65}$ \pm $ \am{0}{15} at
$n_{HI} = 3 \times 10^{19}$~cm$^{-2}$, or roughly 1.5 times the (D$_{26}$)
diameter of the stellar disk.  At $n_{HI} = 1.0\times10^{20}$ cm$^{-2}$, we
measure an \HI\ diameter of \am{8}{15}$ \pm $\am{0}{05}. 

The slight depression in the \HI\ intensity distribution visible near the
optical center of UGC~7321 is possibly due to the mild self-absorption
discussed above.  Such features are frequently seen in \HI\ images of edge-on
spirals.  In addition, small \HI\ filaments appear to be protruding
vertically from the disk to higher latitudes.  These features are discussed
further in Section~\ref{pvz}.

\subsubsection{The {H\,{\sc i}} Warp\protect\label{warp}}
One particularly interesting feature of the total \HI\ intensity map of 
UGC~7321 is the ``integral sign'' warp visible in the \HI\ disk.  The warp
commences near the edge of the stellar disk, reaching a peak amplitude of
$\sim 5''$ on the east side of the disk and $\sim 10''$ on the west (i.e.,
roughly 4\% of the \HI\ disk radius), before the gas twists back toward the
midplane in the outermost regions.  These features can be seen clearly in
the warp curve shown in Figure~\ref{fig:warpcurve}, where we have plotted
the mean deviation of the \HI\ gas from the disk midplane as a function of
radius, measured by fitting single Gaussians to intensity profiles
perpendicular to the disk.

Warps with features similar to those of UGC~7321 have now been observed in
the \HI\ disk of the Milky Way (Burton~1988) and numerous edge-on spiral
galaxies (Sancisi~1976; Briggs~1990; Garc\'\i a-Ruiz et al.~2002).
Gaseous warps usually originate at the edge of the stellar distribution,
which is to be expected because it would seem easier to pull material out
from the midplane in regions where the self-gravity is lowest, and because
once material becomes displaced from the midplane, dissipational and
dynamical-friction forces will cause it to settle back into equilibrium more
quickly within the stellar disk than outside of it (e.g., Bottema 1996).
However, the outer stellar disk of UGC~7321 is extraordinarily diffuse
(Matthews et al.~1999), implying that its self-gravity should be far less
significant than in the outer regions of a galaxy like the Milky Way.
Therefore, the commencement of the \HI\ warp near the edge of the stellar
disk (with essentially no hint of warping being visible in the stars) seems
somewhat surprising. 

A number of researchers have shown that warps are more commonly seen in
galaxies that have companions (e.g., Reshetnikov \& Combes~1998), implying
that interactions play a role in triggering them.  However, UGC~7321 appears
to be an unusually isolated galaxy, with no close companions detectable in
the optical bands, and no gas-rich companions with $M_{HI} \ga 10^{6}$~\msun
(see Section~\ref{COMPANIONS}).  The nearest optically catalogued neighbors
to UGC~7321 are UGC~7236 (type Im), at $V_{HEL}$=945~\kms\ and a projected
distance of \am{115}{6} (0.34~d$_{10}$~Mpc), and NGC~4204 (type SBdm), at
$V_{HEL}=$861~\kms\ and a projected distance of \am{117}{5} (again
0.34~d$_{10}$~Mpc).  Assuming a typical group speed of 200~\kms\ , the last
encounter of UGC~7321 with one of its neighbors would have taken place at
least $\ga 1.6\times10^{9}$ years ago, even if they were separated only by
their (minimal) projected distance.  Is it plausible that an encounter
with one of these systems could have triggered the warp in UGC~7321?

A consequence of tidal interactions will be to cause heating (thickening) of
galaxy disks (e.g., Reshetnikov \& Combes 1997).  Although the stellar disk
of UGC~7321 is (dynamically) much colder than those of typical spiral
galaxies, Matthews~(2000) has shown that UGC~7321 does show distinct
vertical color gradients and a multi-component stellar disk structure,
consistent with a disk that has been mildly heated.

Using statistical arguments, Reshetnikov \& Combes (1997) suggested that the
typical ``cooling'' time for a disk mildly perturbed by an interaction
should be $\sim$10$^9$~years.  Hofner \& Sparke (1994), who treat warps as
discrete bending modes in a disk embedded in a dark halo potential, have
developed a formalism to estimate the age of a warp.  In UGC~7321, the
rotation curve flattens at about $120''$ from the center (the Hofner-Sparke
$r_{0}$), the warp commences at about $r=150''$ from the center of the disk,
the peak rotational velocity  ($V_{MAX}$) is about 110~\kms, the disk scale
length ($h_{r}$) is $2.5$~kpc and the disk mass (in gas and stars) is
$1.34 M_{HI} + \Upsilon_{\star}L_{B}$.  In this latter relation, the factor
of 1.34 corrects the gas mass for the contribution of helium and
$\Upsilon_{\star}$ is the stellar mass-to-light-ratio.  We assume that the
contributions of molecular and ionized gas are negligible.

Using the models of Bell \& de Jong (2001), we adopt a stellar mass-to-light
ratio $\Upsilon_{\star}$=1.0 based on UGC~7321's $B-R$ color of 0.97.  Then,
equation~7 of Hofner \& Sparke (1994) leads to an estimate of the age of the
warp of $\sim$10$^{9}$ years.  Thus, our estimate does not exclude the
possibility that a mild encounter between UGC~7321 and one of its nearest
catalogued dwarf neighbors could have occurred $\sim 10^9$~years ago and
triggered its warp (as well as some mild dynamical heating of its disk).
Assuming that the timescale for the cooling of the disk is
$\sim$10$^9$~years as suggested by Reshetnikov \& Combes~(1997), such a
scenario would have allowed sufficient time for subsequent star formation
(and possible gas accretion) to re-establish the very thin young stellar
disk that is observed in UGC7321.

Ultimately, self-consistent numerically calculations (including the gas and
the stars, as well as a ``live'' dark matter halo) will be needed in order
to gauge whether the amplitude and line of nodes of the warp of UGC~7321 are
consistent with its triggering by the passage of one of its dwarf neighbors,
and whether a thin, moderate-mass disk like UGC~7321 could have survived
such an encounter relatively unscathed.

\section{The {H\,{\sc i}} Kinematics of UGC~7321\protect\label{kinematics}}
\subsection{The Velocity Field\protect\label{velfield}}
We have computed a map of the first moment of the \HI\ emission
(representing the \HI\ velocity field) of UGC~7321 from our $r = 1$ spectral
image cube using the standard ``cutoff'' technique (Rots \& Shane 1975).  A
2$\sigma$ clipping was applied to the absolute values in the image cube
after spatially smoothing the data with a Gaussian kernel of $FWHM = 33''$
and Hanning-smoothing the data in velocity.  In order to maintain a uniform
cutoff with respect to the noise, these operations were performed on images
uncorrected for primary beam attenuation.  The spatial and velocity
smoothing were used for the purpose of deciding which pixels to discard; the 
moment map was made with the full resolution of the input cube.  Because of
the biases that clipping of any sort can introduce (Bosma~1981), we used the
moment-1 map (shown in Figure~\ref{fig:mom1}) only to obtain a qualitative
impression of the global \HI\ velocity field of UGC~7321.
Overall, the velocity field of UGC~7321 is fairly smooth and normal, although
some minor irregularities are visible.  In the innermost regions, the
isovelocity contours exhibit the characteristic ``V''~shape of a
differentially rotating disk, with only a slight asymmetry in the shape of
these contours near the eastern and western ends of the disk.  The symmetry
of these contours about the midplane shows that the (kinematic) major axis of
UGC~7321 does not appear to undergo any significant twisting or precession.

Although the velocity field of UGC~7321 is fairly regular, a mild asymmetry
between the receeding and the approaching sides is noticeable about half-way
through the extent of the optical disk.  In addition, at galactocentric radii
$r\ga$75$''$, the isovelocity contours begin to exhibit some curvature at the
edges that can be attributed to the warping of the disk.  Finally, at the far
edges of the disk, we see the most pronounced deviations from purely regular
rotation.  On the approaching side, the isovelocity contours exhibit their
greatest distortion as the terminal velocity of the disk is neared, while the
receding side of the disk shows three irregular patches of higher velocity
material superimposed on the constant-velocity outer disk region.

\subsection{Position-Velocity Diagrams}
\subsubsection{The Major Axis P-V Diagram\protect\label{pvr}}
In Figure~\ref{fig:PV} we show position-velocity (P-V) diagrams for UGC~7321
along the major axis of the disk (center panel), as well as at heights of
$\pm15''$, above and below the midplane.  These diagrams were made from
unsmoothed data with no averaging applied along the vertical direction.

The overall shape of the P-V diagrams of UGC~7321 is that of a
``scaled-down'' giant spiral (e.g., M31; Brinks \& Shane 1984); although it
shows a smaller fraction of its emission near the terminal velocity (along
the ``flat'' part of the P-V diagram) than in most giant spirals, and the
peak rotational velocity of UGC~7321 is only about half of that of a system
like M31.  A large range of gas velocities is observed at all positions
across the disk, particularly in the central regions.  While similar
features are commonly seen in the P-V diagrams of more luminous spirals,
many low-mass spiral galaxies with luminosities and rotational velocities
similar to those of UGC~7321 exhibit slowly rising, solid-body rotation
curves with a much narrower spread of velocities at any given radius [e.g.,
IC~2233 (Matthews \& Uson~2002);  NGC~55 (Puche, Carignan, \&
Wainscoat~1991); NGC~4395 (Swaters et al.~1999)]. 

One interesting feature of the major axis P-V profile of UGC~7321 is its
``figure-8'' shape, which is clearly visible in Figure~\ref{fig:PV}.  M31
exhibits a similar signature in its major axis P-V~curve, and Brinks \&
Shane~(1984) demonstrated that this results from the warping and flaring of
its \HI\ disk.  As discussed above, the \HI\ disk of  UGC~7321 is also
warped, and the twisting of its gas disk is likely to be the origin of the
``figure-8'' shape seen in this galaxy.  This can be seen
in Figure~\ref{fig:mom1}, where the isovelocity contours are found to be
predominantly V-shaped in the inner disk regions, but at $r\sim75''$ they
begin to twist near the edges.

It is also interesting to note that the central panel of Figure~\ref{fig:PV}
suggests the presence of a (small) bar or inner arm that extends from about
$+30''$ to $-40''$ and manifests itself mainly in the decrease of the \HI\
intensity at $\sim (-25''$, $+80$~\kms).  Indeed, in a barred galaxy viewed
edge-on, the lack of available orbits for gas near the corotation radius
manifests itself as a perturbation in the major-axis P-V diagram (Kuijken \& 
Merrifield 1995).  In addition, the optical H-band major-axis profile shows
two distinct ``shoulders'' at $\sim \pm 25''$ where the light exceeds that of
a pure exponential disk (Matthews et al.~1999, their Figure~3).  Given its
edge-on orientation it might be difficult to distinguish between a small bar
and an inner arm or even confirm their reality.  However, should further
observations confirm that UGC~7321 is in fact barred, they would challenge
the conventional view that bar formation is strongly suppressed in galaxies
with low surface densities and large dark matter contents (Mihos et
al.~1997).

\subsubsection{The Minor Axis P-V Diagrams\protect\label{pvz}}
In Figure~\ref{fig:ZPV} we display P-V diagrams extracted at five different
cuts parallel to the rotational axis of UGC~7321: along the minor axis
($r=0$), at $r=\pm30''$, and at $r=\pm60''$.  In all of these P-V~profiles,
we identify features at $\ge 2 \sigma$ levels that appear to be filaments
extending out of the plane of UGC~7321 to distances up to $\sim60''$
($\sim$2.9~d$_{10}$~kpc).  In addition to these features, other apparent
asymmetries are visible in the minor axis P-V profiles when the two sides of
the disk are compared.  These trends are reminiscent of features reported in
the giant, edge-on Sbc galaxy NGC~891 by Swaters et al.~(1997), which they
attributed to the presence of an \HI\ halo produced by a galactic fountain.
As discussed by these authors, detailed modelling is required in order to
demonstrate conclusively that the apparent high-latitude extensions are not
simply manifestations of a warp or of flaring of the gas disk.  An analysis
of the high-latitude gas will be reported elsewhere (Matthews and Wood~2002).

\subsection{The Disk Rotation Curve of UGC~7321\protect\label{rotcurve}}
\subsubsection{Derivation of the Rotation Curve}
Commonly used methods of converting P-V data into a true disk rotation curve,
such as fitting tilted ring models to the velocity field, extracting major
axis cuts along the velocity field, or fitting single, intensity-weighted
Gaussians to one-dimensional P-V slices, break down for edge-on and nearly
edge-on galaxies as a consequence of projection effects.  Indeed, if applied
to highly inclined galaxies, these methods will result in a rotation curve
which is artificially shallow in the inner regions (Sancisi \& Allen 1979).
Therefore, we have used a technique that was first introduced to Galactic
rotation studies by Shane \& Bieger-Smith~(1966) which involves the
determination of the terminal velocity at a given position, $v_{t}$, from
one-dimensional P-V cuts at various radii using an ``equivalent rectangular
measure'' based on the location of the extreme velocity features:

\begin{equation}
v_{t}=v_{m} + \frac{1}{F_{m}}\sum^{v_{+}}_{v_{m}}F(v)\Delta v
-\frac{\Delta v}{2} -\frac{\sqrt{2\pi}}{2}\sigma,
\end{equation}

\noindent where $F_{m}$ is the peak flux density in the profile, $v_{m}$ is
its corresponding velocity, $\Delta v$ is the channel separation, and
$v_{+}$ is the velocity at which the wing of the extreme velocity edge of
the line profile drops to zero.  The last term in the equation is a
correction that accounts for gas turbulence and was introduced by Burton
\& Gordon~(1978) who also noted that this method is rather insensitive to
noise.  We make the further approximation that $\sigma$ is everywhere equal
to 7~\kms---the average of the range of $\sigma\sim $(5 -- 9)~\kms\ observed
in normal spiral and quiescent dwarf galaxies (van der Kruit \& Shostak~1984;
Lo et al.~1993; Olling~1996) and which we also find to be 
consistent with our data based on Gaussian fits to the extreme
velocity envelopes of our one-dimensional P-V cuts.
We have ignored any additional corrections for asymmetric drift in UGC~7321,
as such corrections for the gaseous components of disks are typically on the
order of only a few \kms---i.e. small compared with the uncertainties of
applying such corrections (de~Blok \& Bosma~2002).

The resulting rotation curve is shown in Figure~\ref{fig:rotcurve}.  We have
indicated on this figure the location of the edges of the stellar disk.
The statistical errors of these measurements have a median value of
0.7~\kms , with only two points at the end of the receding side and three
points at the end of the approaching side having significantly larger
statistical errors of up to 6~\kms .  However, the uncertainties are likely
to be dominated by systematic errors.  We have compared the values that we
derived using the ``equivalent rectangular measure'' to those obtained from
fitting ``half-Gaussian'' functions to the outer part of the profiles in the
flat region of the rotation curve and estimate systematic errors of order
$\sim 3$~\kms .

We see that the rotation curve of UGC~7321 flattens on both sides of the
disk, reaching a peak rotational velocity of $\sim$110~\kms, but that it
rises rather slowly, not reaching $V_{max}$ until about half-way through the
disk.  This is in contrast to high surface brightness spirals, whose
rotation curves tend to rise rapidly to their asymptotic value, (e.g.,
Casertano \& van~Gorkom~1991).

It is clear from Figure~\ref{fig:rotcurve} that the rotation curve of
UGC~7321 does not exhibit any strong asymmetries or lopsidedness,
particularly in the inner regions.  It does however show a slight
($\sim$10\%) difference between the peak amplitudes of the approaching and
receding sides of the disk, and the rotation curve is observed to extend
about 1~kpc further on the approaching side (see also Figure~\ref{fig:PV}.

Figure~\ref{fig:rotcurve} shows that near the edge of its \HI\ disk, the
rotation of UGC~7321 appears to decline, decreasing by $\sim$9\% from its
peak value.  Evidence for this decline can also be seen in the major axis
P-V plot in Figure~\ref{fig:PV} and the channel maps in
Figure~\ref{fig:channelmaps}. A declining rotation curve was also reported
for the thin, edge-on spiral NGC~4244 by Olling (1996).  Two possible
interpretations of this falling rotation curve in UGC~7321 are that either
the dark matter halo of UGC~7321 is truncated near the edge of the \HI\
disk, or that we are seeing a projection effect caused by the warping of
UGC~7321.  Both interpretations are consistent with the data.

Whenever a slowly rising rotation curve is inferred from \HI\ aperture
synthesis measurements, it is of some concern whether the apparent slow rise
could be an artifact of beam smearing (Begeman~1989).  Indeed, some of the
slowly rising rotation curves reported for LSB galaxies in the literature
have been attributed to this effect (Swaters, Madore, \& Trewhella~2000;
van~den~Bosch et al.~2000).  These rotation curves cannot be used to
constrain the shape of dark matter halos, since the inner region is the most
critical for distinguishing between the cuspy halos predicted by cold dark
matter (CDM) models and dark matter halo models with lower density cores.

In order to gauge the possible effects of beam smearing in the rotation
curve of UGC~7321, we have performed two tests.  Firstly, we produced a new
deconvolved image cube of UGC~7321 using a robustness value of $-1$ as
described in Section~\ref{imagline}, which yielded a spatial resolution of
$\sim 12''$, 25\% narrower than that of our initial ($r = 1$) derivation.
The rotation curve obtained following the steps described above is shown as
a dotted line on Figure~\ref{fig:rotcurve}.  Aside from the uncertainties
caused by the $\sim 40$\%  higher noise level in the higher resolution data,
the agreement between the two rotation curves is excellent.  A second check
is provided by the comparison of the \HI\ rotation curve of UGC~7321 with
the rotation curve derived by Goad \& Roberts~(1981) from optical
(H$\alpha$) emission line spectroscopy (shown with `+' signs on
Figure~\ref{fig:rotcurve}).  The optical data have much higher spatial
resolution than the \HI\ data and hence do not suffer from beam smearing.
Given the uncertainties introduced by the more irregular distribution of the
\HII\ regions compared with that of the \HI\ gas, the agreement between the
\HI\ and H$\alpha$ rotation curves is quite good, even in the innermost
regions.  Indeed, maximum deviations are $\la$5~\kms (our spectral
resolution).  As shown by Matthews \& Wood~(2001), the effects of internal
extinction on the optical rotation curve should be negligible in UGC~7321.
We conclude that the effects of beam smearing are insignificant in our
derived \HI\ rotation curve of UGC~7321.

\subsubsection{The Dynamical Mass of UGC~7321\protect\label{dynmass}}
Assuming a spherical mass model for UGC~7321, the dynamical mass of UGC~7321
within radius $r$ is simply $M_{DYN}(r)=V^{2}(r) r G^{-1}$.  Therefore, the
mass contained within the radius of the last measured point of the rotation
curve (taken to correspond to $r = 11.5$~d$_{10}$~kpc, $V = 105$~\kms), is
$M_{DYN} \sim 3.2 \times 10^{10}$~\msun\ d$_{10}$.  This implies ratios of
the total mass to the galaxy's $B$-band luminosity and \HI\ mass,
respectively, of $M_{DYN}/L_{B} = 29$ d$^{-1}_{10}$ and
$M_{DYN}/M_{HI} = 31$~d$^{-1}_{10}$.  These ratios are 7--8 times larger and
3~times larger, respectively, than the median values found by Roberts \&
Haynes~(1994) for Scd--Sd galaxies, indicating a large dark matter content
in UGC~7321.  Assuming a stellar mass-to-light ratio in the range
$\Upsilon_{\star}=1.0$ -- 1.4, consistent with the $B - R$ color of
 UGC~7321, and a correction to the \HI\ mass of 1.34 for the contribution of
helium, the ratio of the dynamical mass to the visible mass in UGC~7321 is
in the range (11--13) d$^{-1}_{10}$, which is again rather large.

We have begun detailed modelling of the underlying distribution of dark
matter in UGC~7321.  Preliminary results indicate a slight preference for a
pseudo-isothermal sphere over the cuspy distributions suggested by CDM
numerical simulations (Navarro, Frenk, \& White~1996).  In all cases,
UGC~7321 is highly dark-matter dominated at all radii beyond
$\sim 1.5$~d$_{10}$~kpc.

\subsection{The Deprojected Radial {H\,{\sc i}} Distribution of 
UGC~7321\protect\label{radialHI}}
Because of the high inclination of UGC~7321, the column densities inferred
directly from the total-intensity map (Figure~\ref{fig:totintmap}) represent
only projected values rather than the true gas surface densities in the disk.
Deprojected gas surface densities are of interest for assessing the
star-formation efficiency of UGC~7321 and for comparing its \HI\ distribution
to that of other less inclined late-type and LSB spirals.

In order to estimate the deprojected \HI\ surface density in UGC~7321 as a
function of radius, we have followed an iterative modelling procedure using
the formalism developed by Irwin and her collaborators (the CUBIT software,
Irwin \& Seaquist 1991; Irwin 1994).  This is a non-linear least-squares
fitting program that operates on a three-dimensional (\mbox{H\,{\sc i}})
spectral-line image cube and is capable of simultaneously determining the
underlying density distribution and velocity field.  We adopted the observed
rotation curve for UGC~7321 (see Section~\ref{rotcurve}) and used it to
determine the density distribution. 

From the fits, we found that we were unable to reproduce the \HI\ density
distribution of UGC~7321 with a single-component disk model either Gaussian
or exponential in $r$ and $z$.  Neglecting the warping and flaring of the
gas, we were able to find a reasonable fit to the data using a two-component
model:  (1) a ring of Gaussian cross-section with a central hole of radius
\as{30}{0}, a scale-height ($\sigma$) of \as{7}{7}, a scale-length of
\as{93}{9}, and peak density of 0.19 atoms cm$^{-3}$; (2) a Gaussian with
scale-height of \as{5}{4}, a scale-length of \as{123}{0}, and a central
density of 0.06 atoms cm$^{-3}$.  Fixing these parameters, we used CUBIT to
compute a face-on rendition of the galaxy and ``observed'' the intensity
distribution along its major axis.  Subsequently, we used this model profile
in order to scale the observed major axis intensity profile of UGC~7321 to a
deprojected (face-on) value which is shown in Figure~\ref{fig:majorint}.
This profile is in good agreement with the one derived for UGC~7321 by
Garc\'\i a-Ruiz et al.~(2002) from Westerbork data using an Abel inversion
technique.

Figure~\ref{fig:majorint} shows that the peak \HI\ surface density in
UGC~7321 is only $\sim$~5.8~\msun~pc$^{-2}$, somewhat smaller than the mean
value of 10~\msun~pc$^{-2}$ found in normal surface brightness Sd galaxies
by Cayatte et al.~(1994) or Broeils \& van~Woerden~(1994).  Moreover, it
appears that at all radii, the gas surface density in UGC~7321 lies below
that required for efficient star formation based on the dynamical criterion
of Kennicutt~(1989).  While this star formation criterion should only be
considered an approximation (UGC~7321 is clearly forming some stars), this
result is nonetheless consistent with the very low star formation rate
implied by its low (optical) surface brightness disk, weak CO emission, and
low radio continuum and far-infrared fluxes.  Although the gas surface
densities observed in UGC~7321 are very low compared with those of normal
spirals, they are however rather similar to the typical (subcritical) \HI\
surface densities seen in other late-type, LSB spiral galaxies (van der
Hulst et al.~1993).  The deprojected functions indicate a drop in the radial
\HI\ surface density distribution of UGC~7321 at $r\approx$12~kpc; a step of
only one synthesized beam (0.8~kpc) leads to a drop from a deprojected
surface density of $n_{HI} \sim 2.3 \times 10^{19}$~atoms~cm$^{-2}$
($\sim~0.2$~\msun~pc$^{-2}$) to $\sim 6 \times 10^{18}$~atoms~cm$^{-2}$
($\sim~0.05$~\msun~pc$^{-2}$, $\sim 2 \sigma$).

\section{Search for {H\,{\sc i}} Companions Near
UGC~7321\protect\label{COMPANIONS}}
We have searched for companions to UGC~7321 within a $25' \times 25'$ field
and over the full velocity range covered by our observations.  Visual
inspection of the image cube shows no recognizable companions.  In addition,
we have done a ``matched-filter'' automated search using a set of Gaussian
functions in frequency as our kernel (Uson, Bagri, \& Cornwell~1991).  For
$\Delta \nu$ in the range (24.4--244)~kHz (1--10~channels) we search for an
expected signal of the form:

\begin{equation}
\langle {\cal A} ( \nu_\circ ) \rangle = { 1 \over N }
\int_{\nu_1}^{\nu_2} S( \nu - \nu_\circ ) {\cal H}_\Delta ( \nu ) d \nu
\end{equation}

{\parindent0pt where} the kernel has the form

\begin{equation}
{\cal H}_\Delta ( \nu ) = \exp [ - 4 \ln 2 ( \nu / \Delta \nu )^2 ],
\end{equation}

{\parindent0pt where} $\nu_1$ and $\nu_2$ are the edges of the observed band,
and the norm $N$ is

\begin{equation}
N = \int_{\nu_1}^{\nu_2} {\cal H}_\Delta ^2 ( \nu ) d \nu .
\end{equation}

The noise is weighted by the same kernel and is approximately equal to

\begin{equation}
\langle \sigma_{\nu_\circ} \rangle \approx \overline{ \sigma_i } \;
N^{-1/2} ,
\end{equation}

{\parindent0pt which} is the value that corresponds to the case in which all
image planes have similar noise.

The method is surprisingly efficient at finding signals even if their
intrinsic shape is not Gaussian.  Indeed, it is a good discriminant of faint
extensions to the disk of UGC~7321 as they fade into the noise.  When we
exclude the galaxy delineated as discussed in Section~\ref{global}, a
search done with a kernel of width equal to one channel yields a set of
amplitudes, {$\cal A$}, which are nearly Gaussian distributed, with tails
extending up to $\pm 5.5 \sigma$.  This is reasonably close to what is
expected from statistics only as the image cube has
$1.9 \times 10^6$~degrees of freedom (so that tails up to $\pm 5.2 \sigma$
are expected if the noise is Gaussian distributed).  In addition, spurious
``absorption'' at the $6.5 \sigma$ level is found against the source
J1217+2239 discussed in Section~\ref{cont}.  We have traced this systematic
to small positional offsets between the three image cubes (corresponding to
the three observational settings) that we have combined to generate the
125~channel image cube.  These offsets are a consequence of residual phase
noise left behind in the calibration.

Because companions are likely to show \HI\ emission with width of up to
$\sim$40~\kms, we have repeated our search with kernels of width
2--10~channels.  Even though the smoothing introduced by the kernel decreases
the number of degrees of freedom, the algorithm ``finds'' somewhat more
extended distributions with tails that reach $\pm 6 \sigma$, especially when
using kernels with a width of a few (3--7) channels (in addition, the line of
sight to J1217+2239 again shows spurious signals due to the image jitter just
described).  This is likely due to a lack of independence between neighboring
channels in excess of the $\sim$4\% expected from Hanning-smoothing, which
could be introduced by low-level errors in the VLA backend electronics, as
well as by errors in the correction of the spectral bandpass during the data
reduction due to insufficient sensitivity (see also Section~\ref{imagline}).
We have verified that these tails have not been introduced because of our
decision to image the line emission without prior subtraction of the
continuum emission (Section~\ref{imagline}) by reprocessing the data after
performing the customary visibility-based subtraction of the continuum
emission (Cornwell et al.~1992).  The resulting spectral images agree with
the ones described above within the noise;  although they have a higher
``zero-point'' uncertainty due to the small number of line-free channels in
IF1.  In addition, the rms noise is not uniform, being significantly lower
for the four line-free channels at each end that are used to define the
continuum.

In addition, the matched filter finds a ``signal'' at a level of
$\sim$\,$7.4\sigma$~$(1.23 \pm 0.17$~mJy); which appears in a single
synthesized-beam area with a width of about six channels.  We are unable to
provide any corroborating evidence that would lead us to believe that this
small blip represents real \HI\ emission.  Moreover, given that the tail of
negative values of {$\cal A$} reaches down to $\sim$\,$-6 \sigma$ for this
kernel, we prefer to treat it at this time  as spurious excess noise.  We
have added three standard deviations to its fiducial value and corrected it
for the attenuation of the primary beam of the VLA antennas in order to
derive a robust upper-limit to the \HI\ mass of the fiducial companions to
UGC~7321 located within a radius of 12~minutes of arc (36~d$_{10}$~kpc) of
\mhi\ $< 2.2 \times 10^6$~d$_{10}^2$~\msun.

Therefore, we find no close \HI\ companion that might be responsible for the
warp of UGC~7321 (Section~\ref{warp}).  Even allowing for a significant dark
matter component in any such companions to UGC~7321, their total masses would
still be below 1\% of the inferred total mass of UGC~7321
(Section~\ref{dynmass}).  Such a mass is likely to be too small to excite a
warp via tidal forcing (Hunter \& Toomre 1969; Kuijken \&
Garc\'\i a-Ruiz~2001), thus requiring either the warp of UGC~7321 to be quite
old (in order to have allowed the perturber sufficient time move outside our
search region), or that the warp was excited by some other mechanism.

\section{Summary}
We have presented high-sensitivity \HI\ images of the isolated,
edge-on, `superthin' spiral galaxy UGC~7321 with a resolution of 24.4~kHz
and a total bandwidth of 3.052~MHz (125~channels).  In the optical, UGC~7321
exhibits a diffuse stellar disk with little dust obscuration, an extremely
small global scale height, and no discernible spheroid component.  These
features indicate that UGC~7321 is a late-type, LSB spiral seen at high
inclination.  UGC~7321 thus allows us to explore both the radial and vertical
structure of the ISM of an ordinary LSB galaxy whose disk has not been
recently perturbed.

UGC~7321 shows similar global \HI\ characteristics to other late-type, LSB
spirals:  It has a gas-rich disk with
\mhi = $(1.06 \pm 0.01) \times 10^{9}$~d$^{2}_{10}$~\msun, and an inferred
ratio of \dark~=~1.0(in solar units), which is at the high end of the observed
values for spiral galaxies.  Assuming a stellar mass-to-light ratio similar to
the sun (based on its global $B-R$ color) UGC~7321 contains a similar fraction
of its visible mass in gas as in stars, suggesting that it has been a
relatively inefficent star former.  UGC~7321 is undetected in the 21-cm
radio continuum ($F_{CONT} = 0.41 \pm 0.25$mJy), which is consistent with its
status as a low star formation rate system.  Assuming a spherical mass
distribution, the ratio of the dynamical mass to the total visible mass
(total gas+stars) is $\sim 12$, implying that UGC~7321 is a highly
dark-matter dominated galaxy.

To first order, the total-\HI\ distribution of UGC~7321 is regular and
symmetric with a slight depression near its center, and extends to
approximately 1.5~times the stellar radius.  The deprojected \HI\ surface
density is low (peak value $\sim5.8~M_{\odot}$ pc $^{-2}$), which is
consistent with other LSB spirals.  The outer \HI\ disk of UGC~7321 displays
mild flaring in addition to an ``integral sign'' warp that commences near
the edge of the stellar disk.  The warped material bends back toward the
equatorial plane in the outermost regions.  The origin of this warp is still
not understood, as UGC~7321 appears to be an isolated galaxy.  The nearest
known companions to UGC~7321 are two dwarf galaxies, both with projected
distances of 0.34 d$_{10}$~Mpc and $\Delta V_{H} > 450$~\kms which would
have had their last encounter with UGC~7321 $\ga 1.6\times10^{9}$~years ago.
In addition, the observations described in this paper rule out the presence
of any close gas-rich companions with \HI\ masses greater than
2.2$\times10^{6}$\msun\ ($10 \sigma$) within $12'$ (36~d$_{10}$~kpc).

The \HI\ disk of UGC~7321 appears thicker and more complex than what a
single, cold \HI\ layer confined to the midplane would display.  However,
the projection effects expected from warping and flaring of the gas layer
have to be modelled in order to untangle the three-dimensional structure of
the galaxy, which is complicated by its near-edge-on orientation.  In
addition, we detect filaments of gas extending to distances of up to
$\sim$60$''$ (2.9~d$_{10}$~kpc) from the mid-plane in the total intensity
\HI\ images and in the position-velocity cuts along the minor axis of the
disk, which suggests the presence of high-latitude gas.

The position-velocity plot along the major axis of UGC~7321 displays a
``figure-8'' pattern that we interpret as a manifestation of the warping
of the \HI\ disk.  In addition, the diagram reveals the presence of a small
bar or inner arm.  Overall UGC~7321 has a rather regular \HI\ velocity field
with only minor perturbations.  Similarly, the \HI\ rotation curve of
UGC~7321 exhibits only a mild asymmetry about the two sides of the disk.  The
rotation curve rises rather slowly and linearly throughout the stellar disk,
indicating a low central matter density; this effect cannot be attributed to
the finite spatial resolution of our observations.  A peak rotational
velocity of $\sim$110~\kms\ is reached near 0.9 optical radii, and a slight
decline in the rotation curve occurs in the last few measured points.  Two
possible interpretations of this decline are that the dark matter halo of
UGC~7321 is truncated near the edge of the \HI\ disk or that we are seeing a
projection effect due to the warping of the \HI\ layer.
\acknowledgements{We are grateful to Eric Greisen for discussions on
calibration and imaging issues as well as for updating several of the
AIPS tasks needed for this work.  We have benefited from  valuable
discussions with John Hibbard, Michael Rupen, Mort Roberts, Dave Hogg, Harvey
Liszt, Butler Burton, Jacqueline van~Gorkom, Michael Pohlen, Jim Condon, Bill
Cotton and Dick Thompson.  LDM was supported by a Jansky Fellowship at the
NRAO and is currently supported by a Clay Fellowship at the
Harvard-Smithsonian Center for Astrophysics.  This research has made use of:
the NRAO-AIPS software package, the NRAO-NVSS survey; the NASA/IPAC Infrared
Science Archive and the NASA/IPAC Extragalactic Database (NED) which are
operated by the Jet Propulsion Laboratory, California Institute of
Technology, under contract with the National Aeronautics and Space
Administration; and the Digitized Sky Surveys (DSS), which were produced at
the Space Telescope Science Institute under U. S. Government grant NAG
W-2166.}

\clearpage
\begin{deluxetable}{lll}
\tablenum{1}
\tablecaption{Visible and Infrared Properties of UGC~7321\tablenotemark{a}}
\tablehead{ Parameter & Value }
\startdata

$\alpha$ (J2000.0) &  12 17 33.8  \nl
$\delta$ (J2000.0) &  +22 32 25  \nl
\nl
Hubble type &  Sd~IV \nl
Distance (Mpc)\tablenotemark{b} & 7 -- 13 \nl
\nl
P. A. (deg) &  $82 \pm 0.5$  \nl
Inclination (deg) & $88 \pm 1$ \nl
$a/b$ & 10.3 \nl
$D_{B25.5}$ (arcmin) & 5.6 \nl
$D_{B25.5}$ (kpc)\tablenotemark{b} & 16.3~d$_{10}$ \nl
$h_{r,R}$ (kpc)\tablenotemark{b} & $(2.1 \pm 0.2)$~d$_{10}$  \nl
$h_{z,R}(0)$ (pc)\tablenotemark{b} & $(150 \pm 20)$~d$_{10}$ \nl

\nl

$B-R$ & 0.97  \nl
$m_{B}$ & 13.8 \nl
$M_B$\tablenotemark{c} & $-17.1$ \nl
$L_{B} (L_{\odot})$\tablenotemark{b} & 1.1$\times10^{9}$~d$_{10}^2$ \nl
$\mu_{B,i}(0)$ (mag arcsec$^{-2}$)\tablenotemark{d} & $\sim 23.5$ \nl

\nl

$S_{FIR}$ (W~m$^{-2}$) & $\sim 2 \times10^{-14}$  \nl

\enddata

\tablenotetext{a}{Values are from Matthews~2000 (and references therein),
with a new correction for foreground extinction where appropriate
($A_{B}$=0.121 from Schlegel, Finkbeiner, \& Davis~1998; instead of
$A_{B}$=0.040 from Burstein \& Heiles~1984).  The errors in the intrinsic
physical quantities are formal errors; true uncertainties are dominated by
the uncertainty in the adopted distance.}
\tablenotetext{b}{We adopt a distance of 10~Mpc (see text), d$_{10}$ is the
actual distance expressed in units of 10~Mpc.}
\tablenotetext{c}{Corrected for internal and external extinction}
\tablenotetext{c}{Deprojected central surface brightness corrected for
internal and external extinction}
\end{deluxetable}

\clearpage
\begin{deluxetable}{ll}
\tablenum{2}
\tablecaption{Summary of UGC~7321 Observations}
\tablehead{ Parameter & Value }
\startdata
\nl
\multicolumn{2}{c}{Observing Set-Up} \nl
\nl
\tableline
\nl
Array configuration &  C \nl

Baseline range (m) & 30 -- 3385 \nl

Observing dates ($ 2 \times 8$~hours) &  2000 May 26 (Day~1), 30 (Day~2) \nl

Phase center, $\alpha$ (J2000.0) &  12$^{h}$ 17$^{m}$ 33.8$^{s}$ \nl
Phase center, $\delta$ (J2000.0) & +22$^{\circ}$ 32$'$ 25.0$''$ \nl

Total on-source observing time (min) & 740 (720 at elevation $ > 40^\circ$) \nl

Flux calibrator &  1331+305 (3C~286) \nl
Phase calibrator  &  1221+282 \nl

Number of IFs &  4 [2 $\times$ (RR, LL)]\nl

Channel width (after Hanning smoothing, kHz) &  24.4 \nl

Velocity separation of channels (km s$^{-1}$) & $\sim$~5.2 \nl

Number of channels per IF &  63 \nl

Heliocentric center velocity (optical definition, km s$^{-1}$)    &  407.0
(IF pair 1, Days 1 \& 2) \nl
\dots                             &  650.0 (IF pair 2, Day 1) \nl
\dots                             &  300.0 (IF pair 2, Day 2) \nl

Primary beam $FWHM$ (arcmin) &  $\sim 30.6$ \nl

\nl
\tableline

\nl
\multicolumn{2}{c}{Deconvolved Image Characteristics} \nl 
\nl
\tableline

\nl

Robustness parameter (r)\tablenotemark{a} &  1 \nl

Synthesized beam $FWHM$ (arcsec) & $\sim 16.2 \times 15.7$ \nl

Synthesized beam position angle (deg) & $\sim -36$\nl

Linear resolution of synthesized beam (kpc)\tablenotemark{b} &
$\sim$ 0.8~d$_{10}$ \nl

RMS noise per channel (1$\sigma$, mJy~beam$^{-1}$) & 0.36 -- 0.40 \nl

RMS noise in Total-intensity image (1$\sigma$, mJy beam$^{-1}$ km s$^{-1}$)
& $\sim 6$ \nl

Limiting column density per channel (1$\sigma$, atoms cm$^{-2}$) &
(8 -- 9)$\times10^{18}$ \nl

RMS noise in \HI\ mass per channel, M$_{\HI\ }$ ($1 \sigma$,
M$_\odot$)\tablenotemark{b} & (4.5 -- 4.9)$ \times 10^4$~d$_{10}^2$ \nl

\enddata

\tablenotetext{a}{We also used r~$= - 1$~($FWHM = 12.4'' \times 12.0''$,
PA~=~$\sim -48^\circ$) to make higher-resolution spectral images and r~=~0.7
($FWHM = 15.3'' \times 15.0''$, PA~$= -28.4^\circ$) to make the continuum
image (see sections 4.3 and 4.2, respectively.}
\tablenotetext{b}{d$_{10}$ is the distance to UGC~7321 expressed in units of
10~Mpc (the distance to UGC~7321, see Table~1).}

\end{deluxetable}

\clearpage
\begin{deluxetable}{llr}
\tablenum{3}
\tablecaption{21-cm Radio Properties of UGC~7321}
\tablehead{ Parameter & Value }
\startdata
\nl
\multicolumn{2}{c}{Measured Parameters} \nl
\nl
\tableline
\nl
Peak \HI\ column density ($16''$ resolution, atoms~cm$^{-2}$) & $7.3 \times
10^{21}$ \nl

$D_{HI}$ (arcmin)\tablenotemark{a} & $8.15 \pm 0.05$ \nl

$\theta_{HI,b}$($FWHM$, arcsec)\tablenotemark{b} & $16.7 \pm 0.3$ \nl

$(a/b)_{HI}$\tablenotemark{c} & 29 \nl

$\int F_{HI}{\rm d}\nu$ (Jy km s$^{-1}$) & $45.3 \pm 0.5$ \nl

$W_{20}$ (km s$^{-1}$)\tablenotemark{d} & 234.3 \nl

$W_{50}$ (km s$^{-1}$)\tablenotemark{d} & 219.8 \nl

$V_{hel,HI}$ (optical definition, km s$^{-1}$) & 406.8 \nl

$F_{cont}$(21 cm) (mJy) & $0.41 \pm 0.25$ \nl

\nl
\tableline
\nl
\multicolumn{2}{c}{Derived Quantities}\nl
\nl
\tableline
\nl
$M_{HI}$ ($M_{\odot}$)\tablenotemark{e} & $(1.06 \pm 0.01) \times10^{9}$
d$_{10}^2$ \nl

$M_{DYN}$ ($M_{\odot}$)\tablenotemark{f} & 3.2$\times10^{10}$ d$_{10}$ \nl 

$M_{HI}/L_{B}$ ($M_{\odot}/L_{\odot}$) & 1.0  \nl

$M_{DYN}/M_{HI}$ & 31.0 d$_{10}^{-1}$ \nl

$D_{HI}/D_{B25.5}$ & 1.5 \nl

\enddata

\tablenotetext{a}{Measured at a column density of $10^{20}$ cm$^{-2}$.}

\tablenotetext{b}{Projected thickness of \HI\ layer along minor axis measured
at a column density of $10^{20}$~atoms~cm$^{-2}$ and corrected for the
resolution of the synthesized beam.}

\tablenotetext{c}{Axial ratio of \HI\ disk measured at a column density of
$10^{20}$~atoms~cm$^{-2}$ and corrected for the resolution of the synthesized
beam.}

\tablenotetext{d}{$W_P$ is the \HI\ profile width measured at P\% of the peak
value}

\tablenotetext{e}{Assuming the \HI\ is optically thin (but see
Section~5.3.1), d$_{10}$ is the distance to UGC~7321 expressed in units of
10~Mpc (the distance to UGC~7321, see Table~1).}

\tablenotetext{f}{From $M_{DYN} = 2.326 \times 10^{5}$~$r$~$V^{2}(r)$, where
we have taken $r = 11.5$~kpc and $V(r) = 105$~km~s$^{-1}$.}

\clearpage
\end{deluxetable}


\bigskip

\clearpage
\bigskip

\clearpage
\bigskip

\clearpage
\bigskip

\clearpage
\bigskip

\clearpage
\addtocounter{figure}{-1}
\bigskip
\begin{figure*}
\plotfiddle{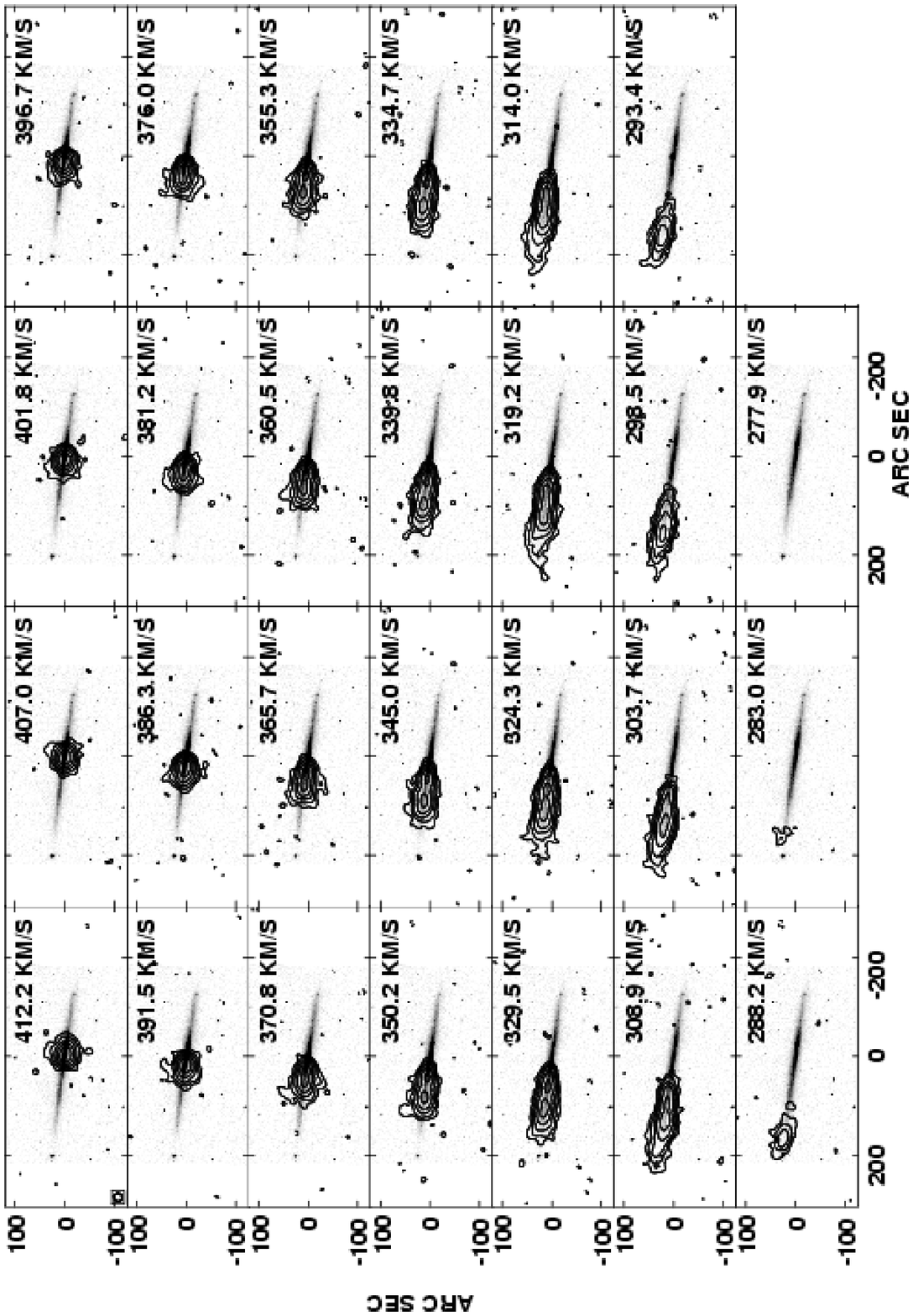}{22cm}{0}{85}{85}{-270}{-10}
\figcaption{(continued)}
\end{figure*}
\clearpage
\bigskip
\begin{figure*}
\plotone{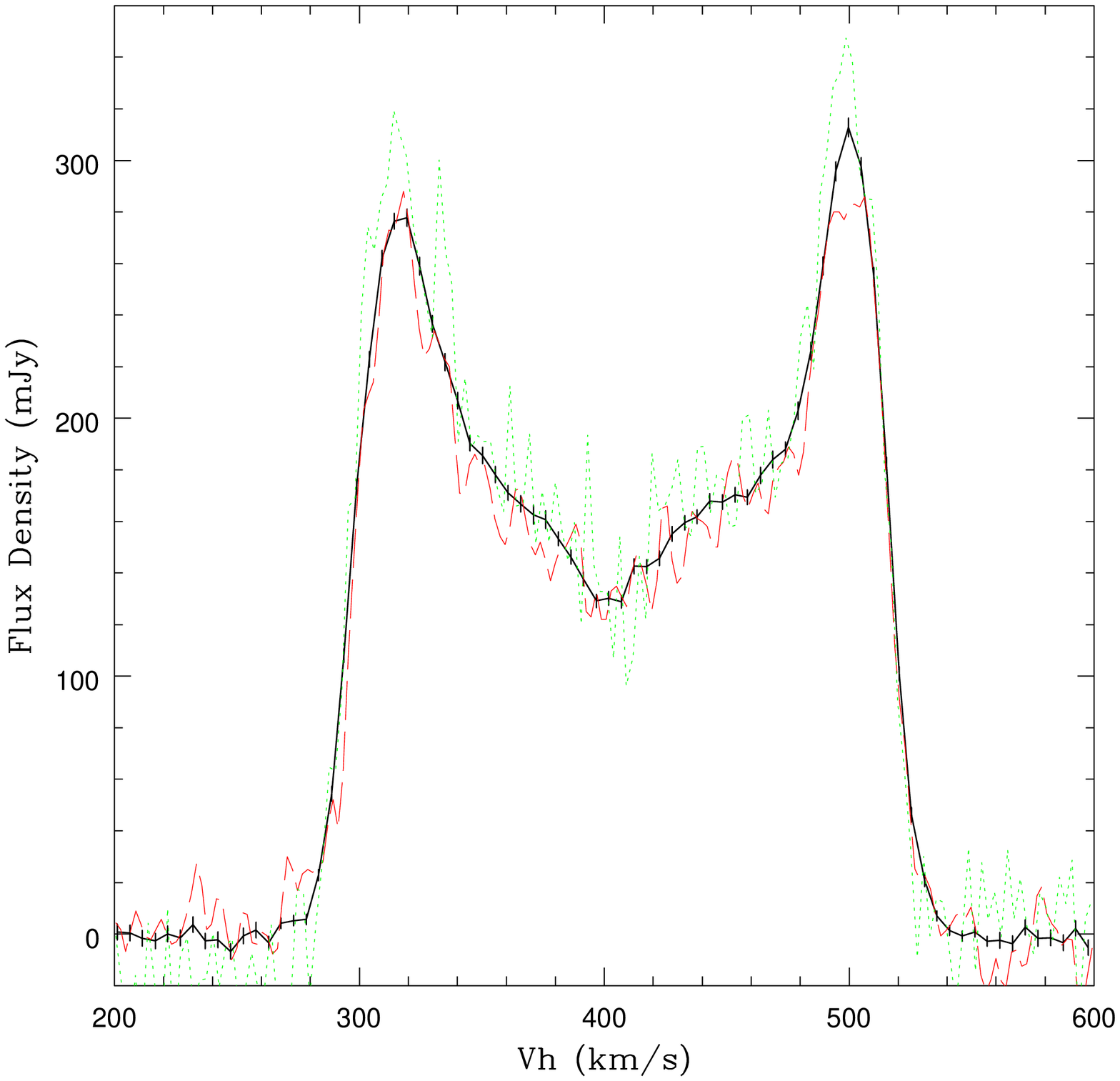}
\figcaption{Global \HI\ profile of UGC~7321 derived from our VLA data
(solid line).  The ($1 \sigma$) error bars are derived from the statistics of
the images and do not include calibration and baseline uncertainties.  Also
shown are the global \HI\ profiles measured with the NRAO 43~m. telescope
(dashes; Haynes et al.~1998) and the \nan\ telescope (dots; Matthews et
al.~1999).
\protect\label{fig:global}}
\end{figure*}

\clearpage
\bigskip
\begin{figure*}
\plotfiddle{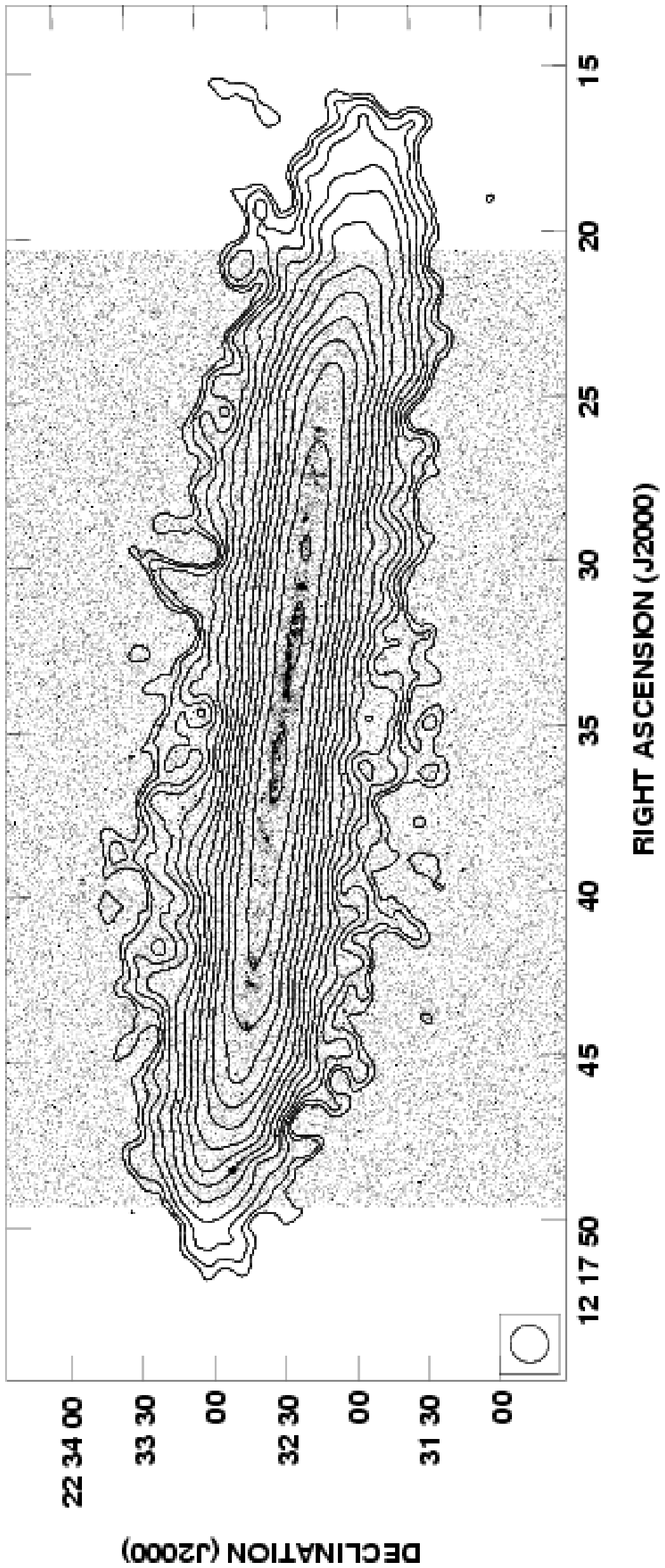}{10cm}{-90}{75}{75}{-300}{300}
\figcaption{\HI\ total intensity contours for UGC~7321, overplotted on an
H$_\alpha+[\NII]$ image of the galaxy from Matthews et al. 1999.  The
synthesized-beam is nearly circular with a FWHM~$\sim16''$.  The contour
levels are (-1 [absent], 1, 1.4, 2, 2.8, 4, 5.6, 8, 11, 16, 22, 32, 44, 64,
88) $\times 18$~Jy~beam$^{-1}$~m~s$^{-1}$.  The \HI\ peaks at
1661~Jy~beam$^{-1}$~m~s$^{-1}$.  Note the slight depression visible near the
center of the disk as well as the ``integral sign'' shape seen in the outer
\HI\ contours, commencing near the edge of the stellar disk.
\protect\label{fig:totintmap}}
\end{figure*}

\clearpage
\bigskip
\begin{figure*}
\plotfiddle{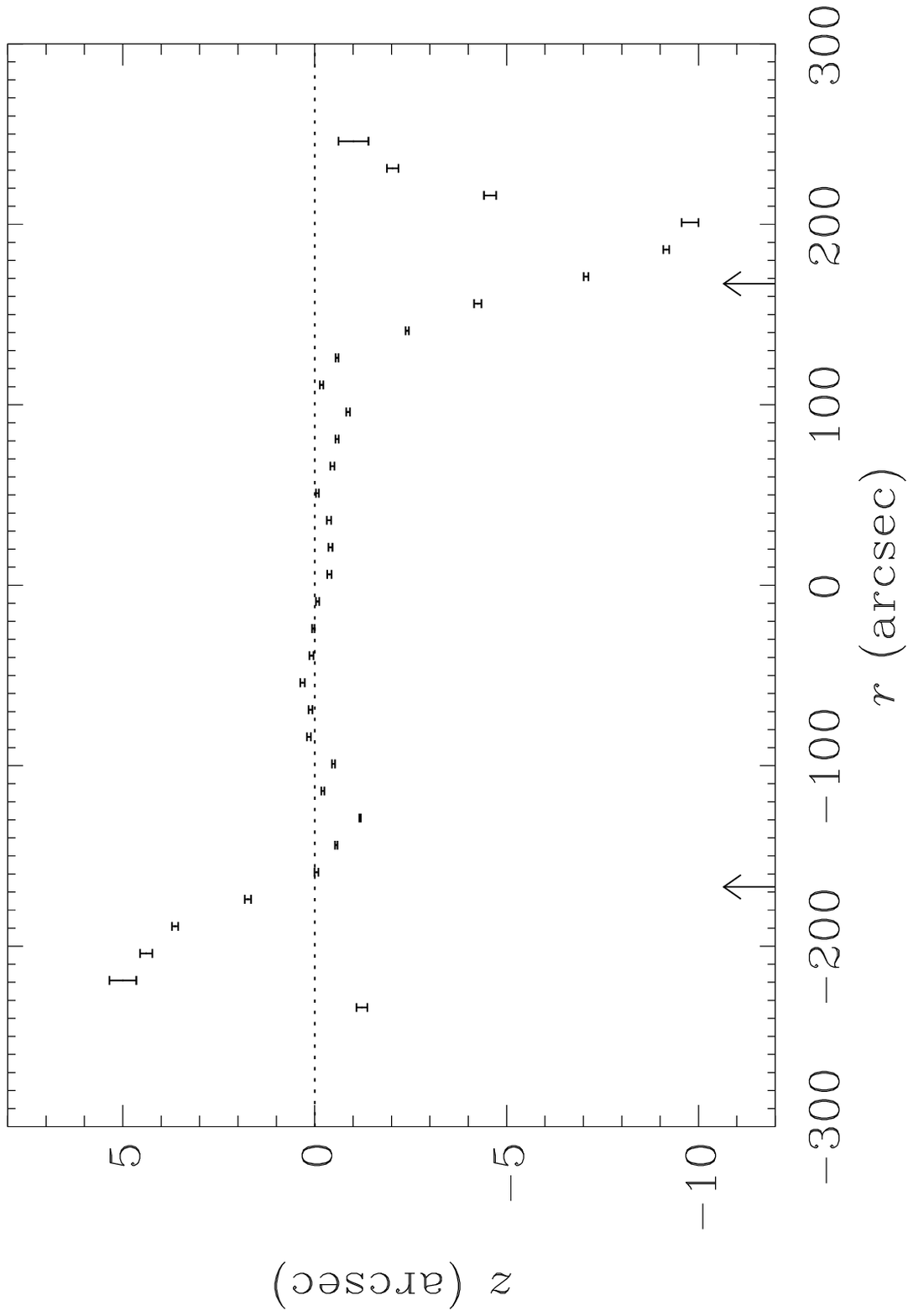}{14cm}{-90}{95}{95}{-400}{500}
\figcaption{\HI\ warp curve derived from UGC~7321 by fitting Gaussians to
\HI\ intensity profiles perpendicular to the major axis at various
locations along the disk. The centroid of
each fitted Gaussian is plotted at each fit location along with the 1$\sigma$
error bars derived from the formal uncertainty in the fit. Note that the
warped material appears to curve back toward the equatorial plane in
the outermost regions of the disk. The warp also shows a slightly
different amplitude on the two sides of the disk.  The arrows show the
location of the observed edges of the stellar disk.
\protect\label{fig:warpcurve}}
\end{figure*}

\clearpage
\bigskip
\begin{figure*}
\plotfiddle{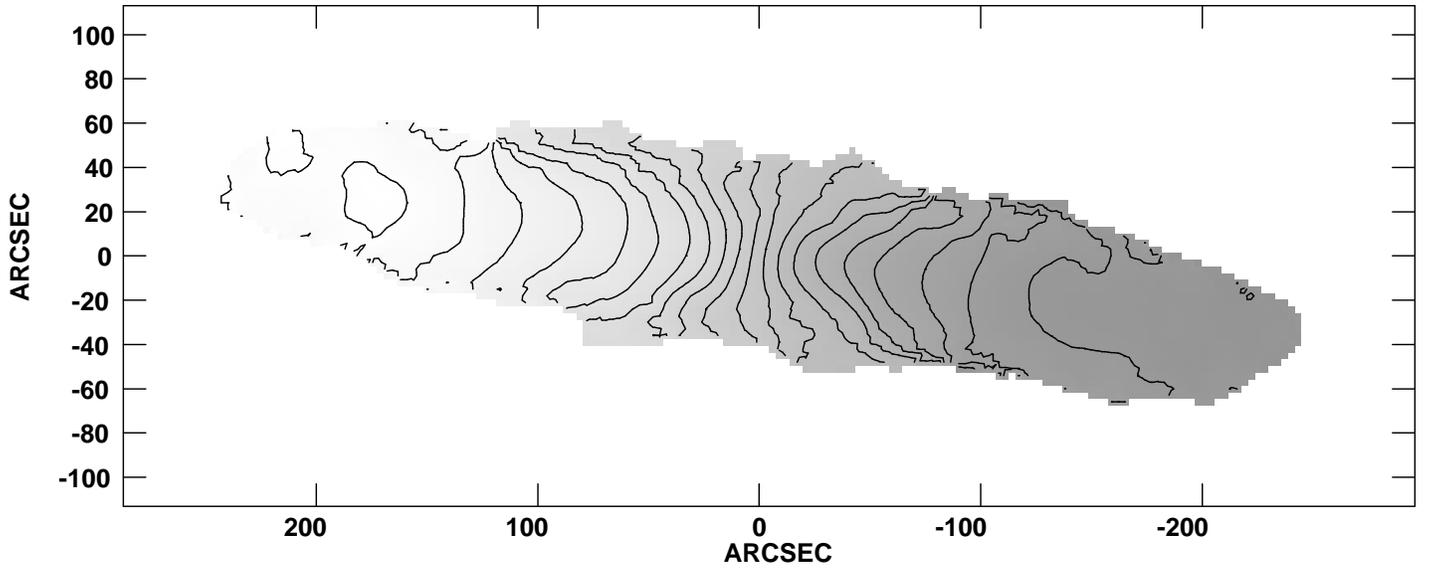}{10cm}{-90}{75}{75}{-300}{400}
\figcaption{\HI\ velocity field for UGC~7321. Isovelocity contours in the
range (303.7--520.7)~\kms\ are shown (every second channel, see
Figure~\ref{fig:channelmaps}), overplotted on a greyscale representation of
the same map.  The grey-scale is linear from 303.7~\kms (lightest) to
520.7~\kms (darkest).
\protect\label{fig:mom1}}
\end{figure*}

\clearpage
\bigskip
\begin{figure*}
\plotone{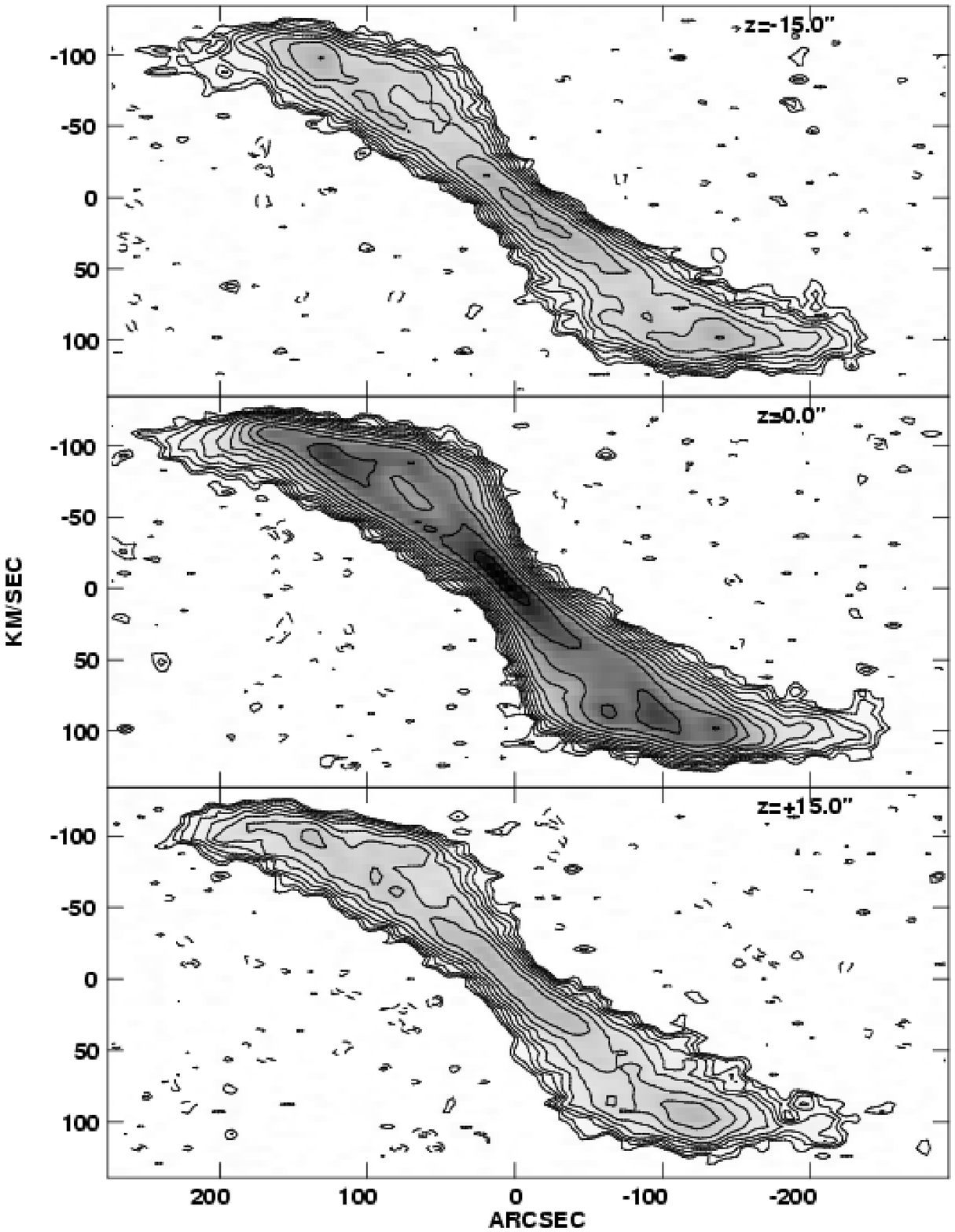}
\figcaption{\HI\ position-velocity (P-V) cuts parallel to the major axis
of UGC~7321.  These figures were derived from images with a spatial
resolution of $\sim16''$ and a velocity resolution of $\sim$5.2~\kms.  The
cuts were extracted along the major axis and at $z = \pm 15''$.  No averaging
was applied along the $z$ direction.  The horizontal axis measures distance
to the minor axis of the galaxy.  The intensity shown is linear in the range
(0--41)~mJy/beam. The contour levels are: (-1.4 [absent], -1, 1, 1.4, 2, 2.8,
4, 5.6, 8, 11, 16, 22, 32, 44) $\times 0.8$~mJy/beam.
\protect\label{fig:PV}}
\end{figure*}

\clearpage
\bigskip
\begin{figure*}
\plotfiddle{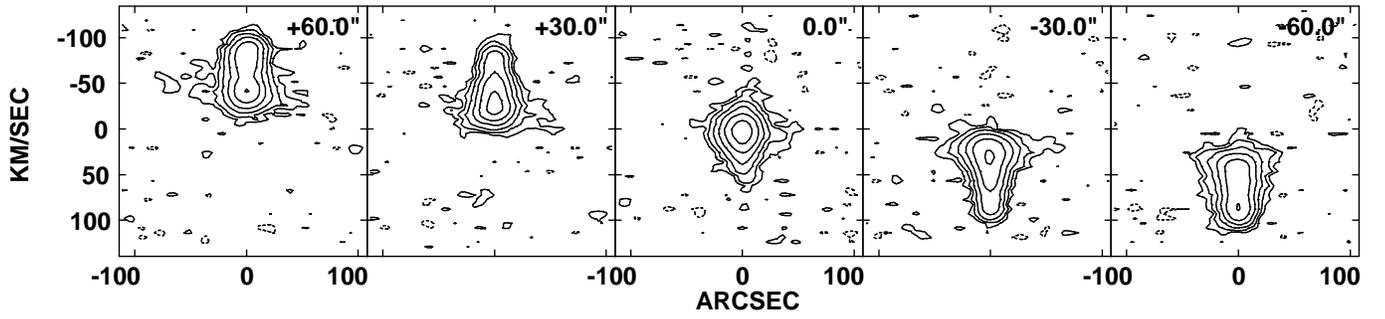}{8cm}{-90}{72}{72}{-280}{350}
\figcaption{Position-velocity (P-V) plots at 5 locations parallel
to the minor axis of UGC~7321.  These figures were derived from images with a
spatial resolution of $\sim16''$ and a velocity resolution of $\sim$5.2~\kms.
The center panel shows the P-V cut along the minor axis; the other panels
show cuts at positions $r = \pm 30''$ and $r = \pm 60''$.  No averaging was
applied along the $r$ direction.  The horizontal axis measures the height
above and below the midplane.  The contour levels are: (-2 [absent], -1, 1,
2, 4, 8, 16, 32, 64) $\times 0.8$~mJy/beam.
\protect\label{fig:ZPV}}
\end{figure*}

\clearpage
\bigskip
\begin{figure*}
\plotone{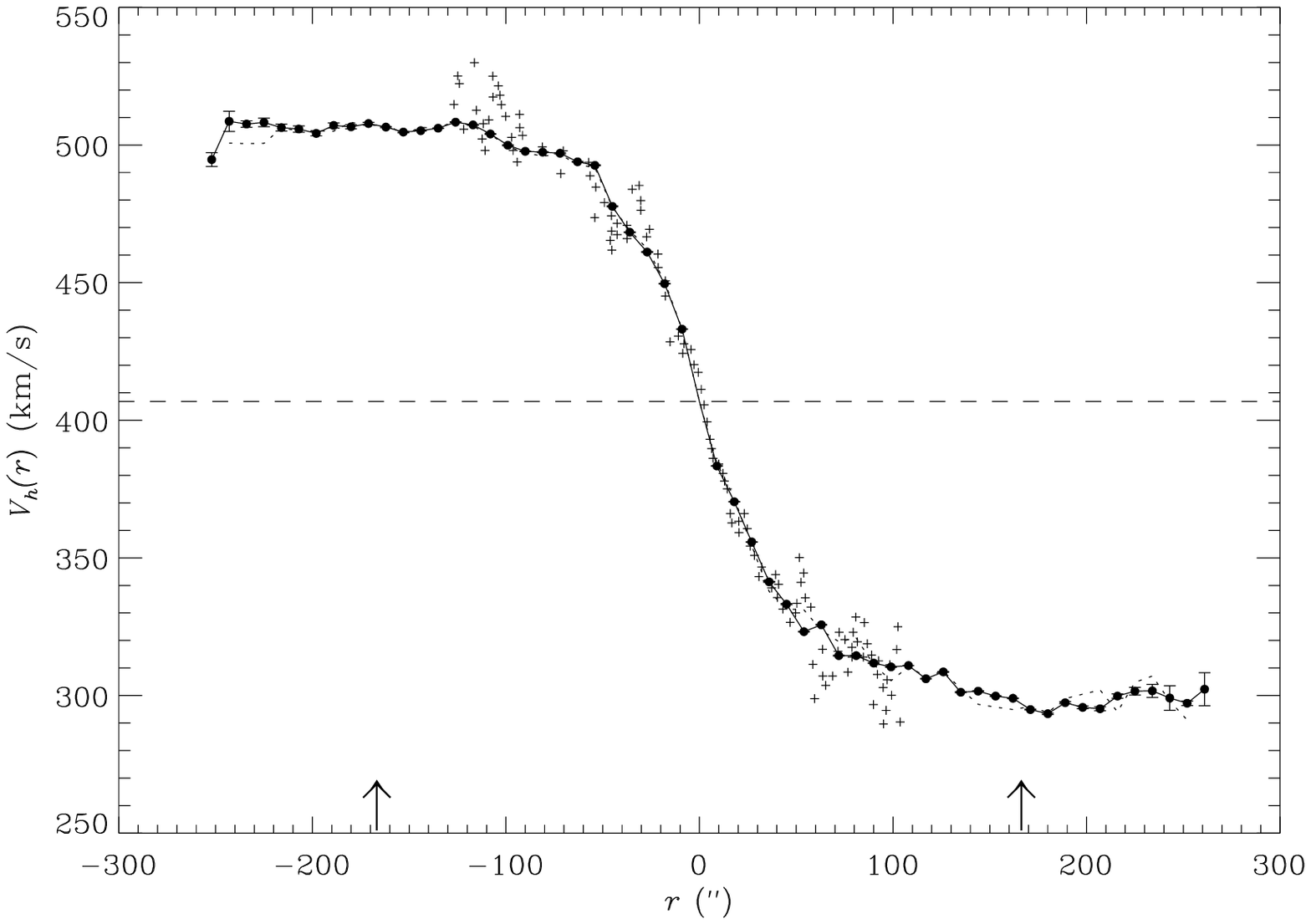}
\figcaption{\HI\ rotation curve for UGC~7321 derived from data with
resolution 16$''$ (solid line and dots) and 12$''$ resolution (dotted line).
The crosses were derived by Goad \& Roberts~(1981) using H$\alpha$
emission-line spectroscopy.  The axes are the distance from the center of the
disk and the heliocentric rotational velocity.  The horizontal dashed line
indicates the systemic velocity inferred from the global \HI\ profile
(Section~\ref{global}).  The location of the observed edges of the stellar
disk are indicated with arrows.
\protect\label{fig:rotcurve}}
\end{figure*}

\clearpage
\bigskip
\begin{figure*}
\plotone{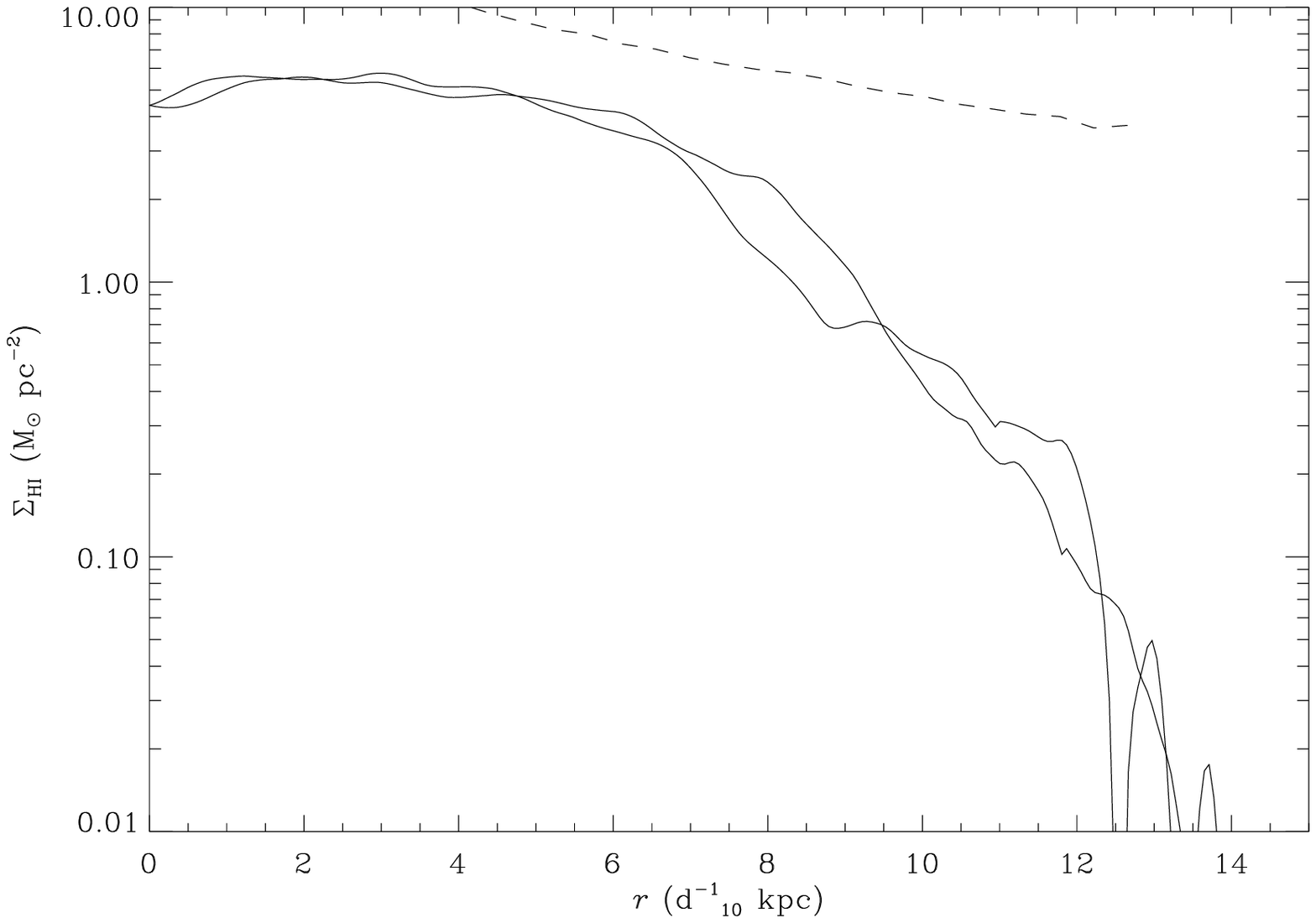}
\figcaption{Deprojected radial \HI\ profile along the major axis of UGC~7321,
folded about $r=0$ (solid lines).  The units are galactocentric radius (in
$d^{-1}_{10}$~kpc) and \HI\ surface density (in
\mbox{${\cal M}_\odot$}~pc$^{-2}$, logarithmic scale).  Overplotted as a
dashed line is the critical \HI\ surface density for star formation according
to the dynamical criterion of Kennicutt~1989 (see text).
\protect\label{fig:majorint}}
\end{figure*}
\clearpage

\end{document}